 \documentclass[twocolumn]{aastex631}

\shorttitle{Light Curves of YZ Ret}
\shortauthors{Hachisu \& Kato}

\begin{document}

\title{A multiwavelength light curve analysis of the classical nova YZ Ret: \\
An extension of the universal decline law to the nebular phase}


\author[0000-0002-0884-7404]{Izumi Hachisu}
\affil{Department of Earth Science and Astronomy, 
College of Arts and Sciences, The University of Tokyo,
3-8-1 Komaba, Meguro-ku, Tokyo 153-8902, Japan} 
\email{izumi.hachisu@outlook.jp}




\author[0000-0002-8522-8033]{Mariko Kato}
\affil{Department of Astronomy, Keio University, 
Hiyoshi, Kouhoku-ku, Yokohama 223-8521, Japan} 

%
%




\begin{abstract}
YZ Ret is the first X-ray flash detected classical nova, and is also well
observed in optical, X-ray, and gamma-ray.  
We propose a comprehensive model that explains the observational properties.
The white dwarf mass is determined to be $\sim 1.33 ~M_\sun$ that reproduces
the multiwavelength light curves of YZ Ret, from optical, X-ray, and to
gamma-ray.  We show that a shock is naturally generated far outside the 
photosphere because winds collide with themselves.  The derived lifetime
of the shock explains some of the temporal variations of emission lines. 
The shocked shell significantly contributes to the optical flux
in the nebular phase.  The decline trend of shell emission in the nebular
phase is close to $\propto t^{-1.75}$ and the same as the universal
decline law of classical novae, where $t$ is the time from the outburst.
\end{abstract}




\section{Introduction}
\label{introduction}

A classical nova is an explosion of hydrogen-rich envelope
on a mass-accreting white dwarf (WD), 
which is triggered by unstable hydrogen shell-burning \citep[see,
e.g.,][for a recent fully self-consistent nova model]{kat22sha}.
The classical nova YZ Ret went into outburst in 2020.
It is characterized by the first detection of an X-ray flash
in classical novae \citep{kon22wa, kat22shb}.
GeV gamma-rays were observed during ten days just after the 
optical maximum, which indicates a formation of a strong shock 
\citep{sok22ll}.  The shocked shell outside the photosphere plays
an essential role in the gamma-ray emission \citep[see, e.g.,][for
a recent review]{cho21ms}.  X-rays are also detected from 10 days after
the outburst \citep{sok22ll}.  YZ Ret was also observed
with longer wavelengths such as optical, infrared, and radio.
In the present paper, we clarify the nature
of YZ Ret by modeling the multiwavelength light curves of YZ Ret, especially,
in optical, X-ray, and gamma-ray light curves in the decay phase of the nova.

In what follows, we describe the characteristic properties of the classical
nova YZ Ret as well as a common evolution of a nova outburst.
We also establish the temporal variations of optical fluxes from 
the shocked shell and show that the flux in the nebular phase follows 
the same trend of $L_{\rm opt, shell} \propto t^{-1.75}$  as that of the
universal decline law of classical nova \citep{hac06kb}.  This extension
of the $L \propto t^{-1.75}$ law to the nebular phase can explain
the decay trends of other classical novae.     

This paper is organized as follows.  First we summarize the multiwavelength
observations of YZ Ret in Section \ref{observational_constrants}.  Section
\ref{light_curve_fitting_yz_ret} describes our light curve fitting with
YZ Ret.  Emissions from a shocked shell are studied separately in
Section \ref{section_emission}.  We propose an optical decline law of
$L_{\rm opt, shell} \propto t^{-1.75}$ in the nebular phase and apply
this law to various novae in Section \ref{t175_law_nebular_phase}. 
Conclusions follow in Section \ref{conclusions}.  In Appendices 
\ref{section_nova_model} and \ref{light_curve_model}, we summarize
our basic nova model and present our light curve model, respectively.

\begin{figure*}
\epsscale{0.85}
\plotone{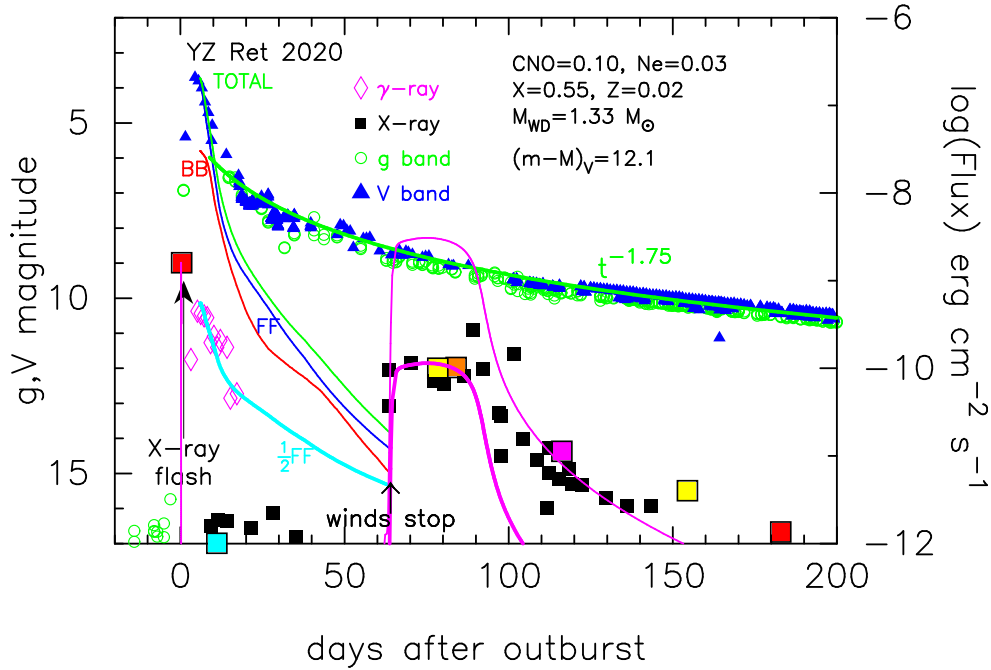}
\caption{
Multiwavelength light curves of the YZ Ret 2020 outburst.
Optical $V$ (filled blue triangles) and $g$ (open green circles) magnitudes
are taken from \citet{kon22wa}.
X-ray fluxes (filled black squares by Swift,
filled red squares by eROSITA,
filled cyan square by NuSTAR, 
filled orange square by NICER,
filled yellow squares by XMM-Newton,
filled magenta square by Chandra).
Gamma-ray fluxes (open magenta diamonds by Fermi/LAT).
The origin of time is set to be $t_0=$ UT 2020 July 7, 16:47:20 
$=$ JD 2,459,038.19954 $=$ MJD 59,037.69954, which marks the first
detection of YZ Ret in outburst.  Solid lines show our best fit model of
a 1.33 $M_\sun$ WD (Ne2).
The red line labeled BB is the $V$ magnitude calculated by a blackbody
approximation of the model photosphere. The blue line (FF) is the
$V$ magnitude from free-free emission of the wind.  The thin
green line (TOTAL) is the sum of BB and FF fluxes.
These $V$ light curves are shown from near optical $V$ maximum to the end
of the wind.  The thick green line labeled $t^{-1.75}$ 
shows an approximated $V$ magnitude model of equation 
(\ref{nebular_emission_brightness}) in the nebular phase.
The thin/thick magenta lines are the soft (0.3--2.0 keV) X-ray
flux calculated by a blackbody approximation of the model photosphere;
the thin line corresponds to unocculted X-ray fluxes same as in the X-ray
flash on $t=0$ day, while the thick one does to the occultation of
the WD surface.  The cyan line labeled ${1 \over 2}$FF
denotes the slope half as fast as that of FF (blue line), which broadly
follows the decline trend of the gamma-ray flux.
See Section \ref{light_curve_fitting_yz_ret} for the detail of fitting.
\label{optical_mass_yz_ret_x55z02o10ne03_no2}}
\end{figure*}


\begin{figure*}
\gridline{\fig{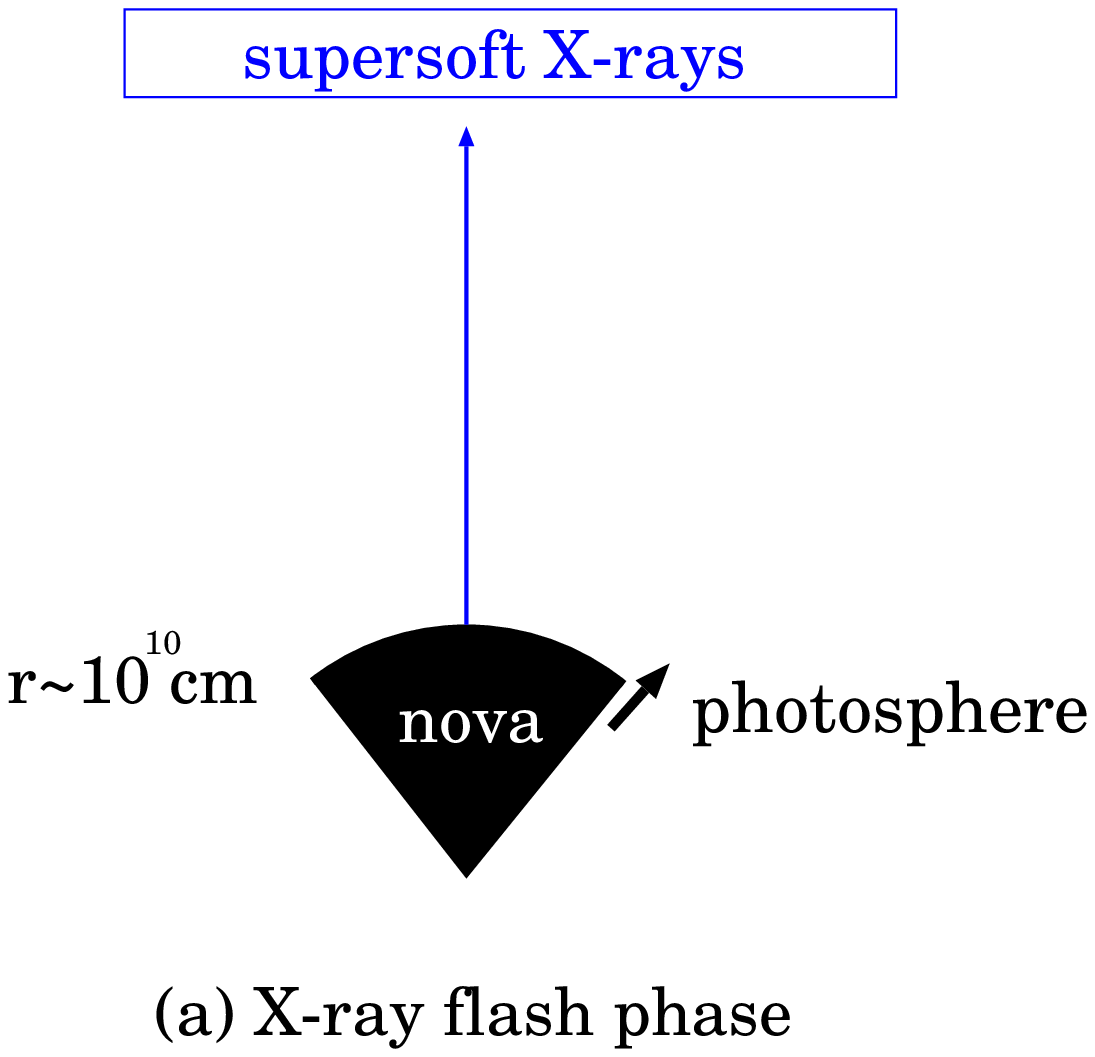}{0.35\textwidth}{}
          \fig{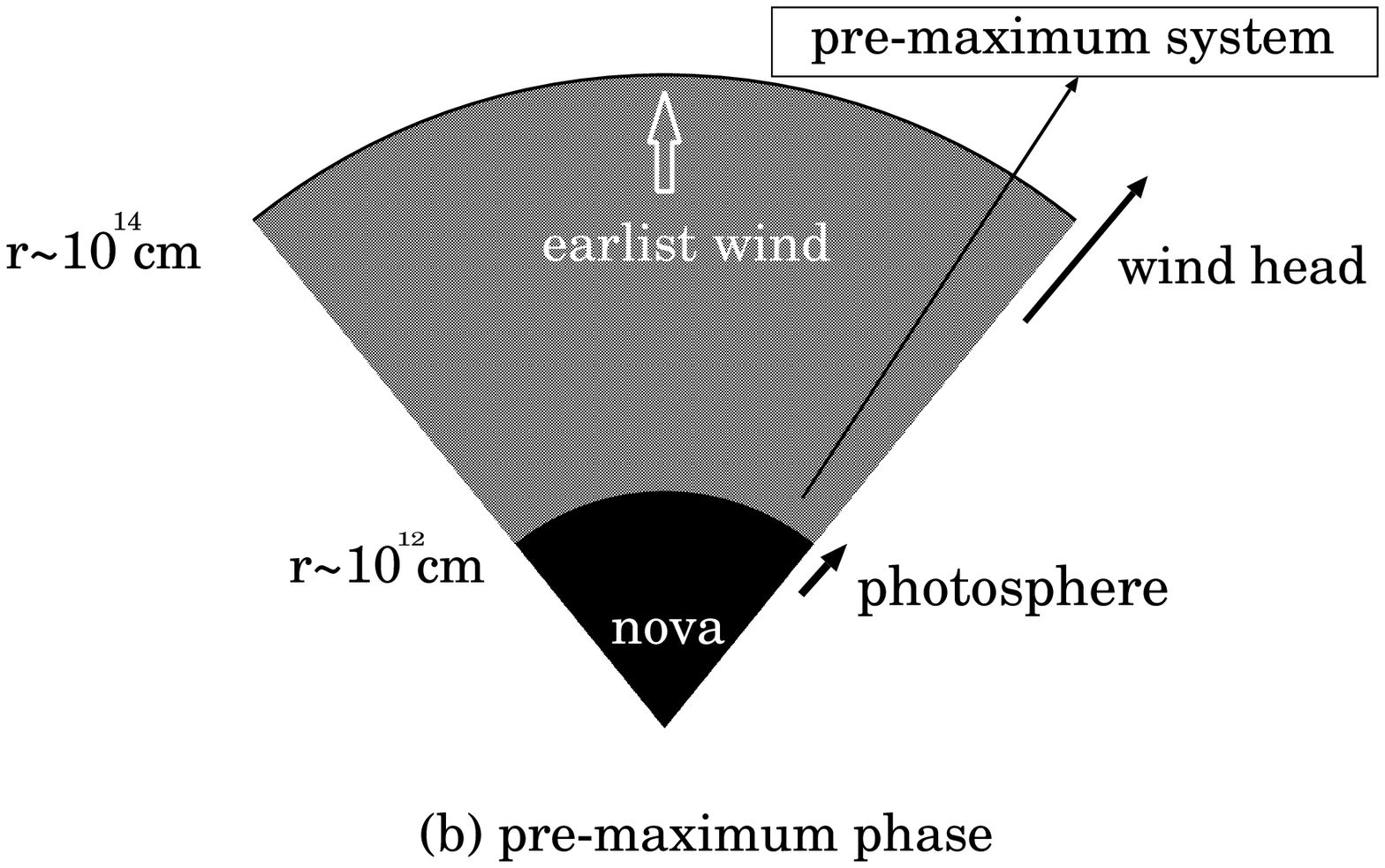}{0.55\textwidth}{}
          }
\gridline{
          \fig{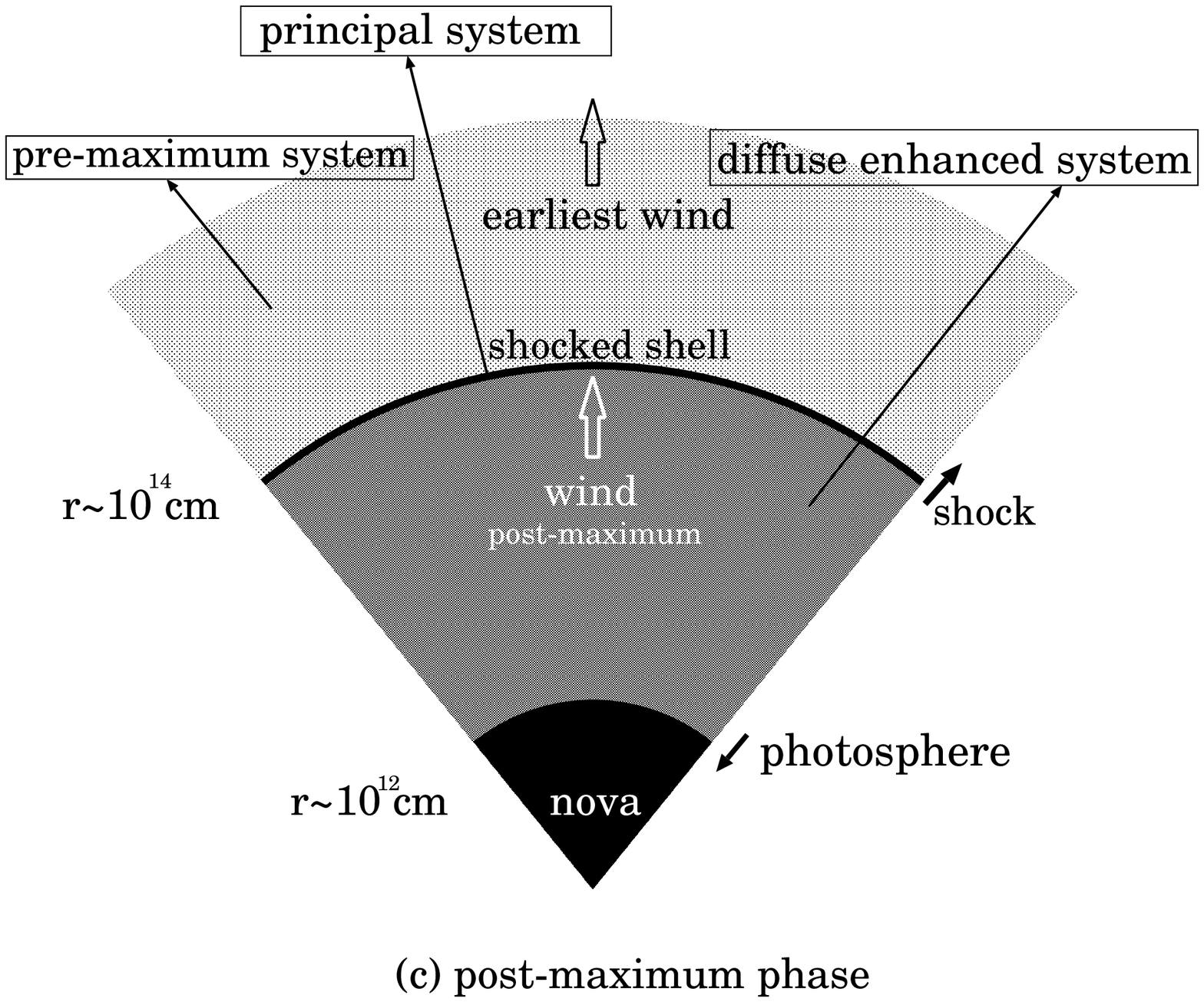}{0.49\textwidth}{}
          \fig{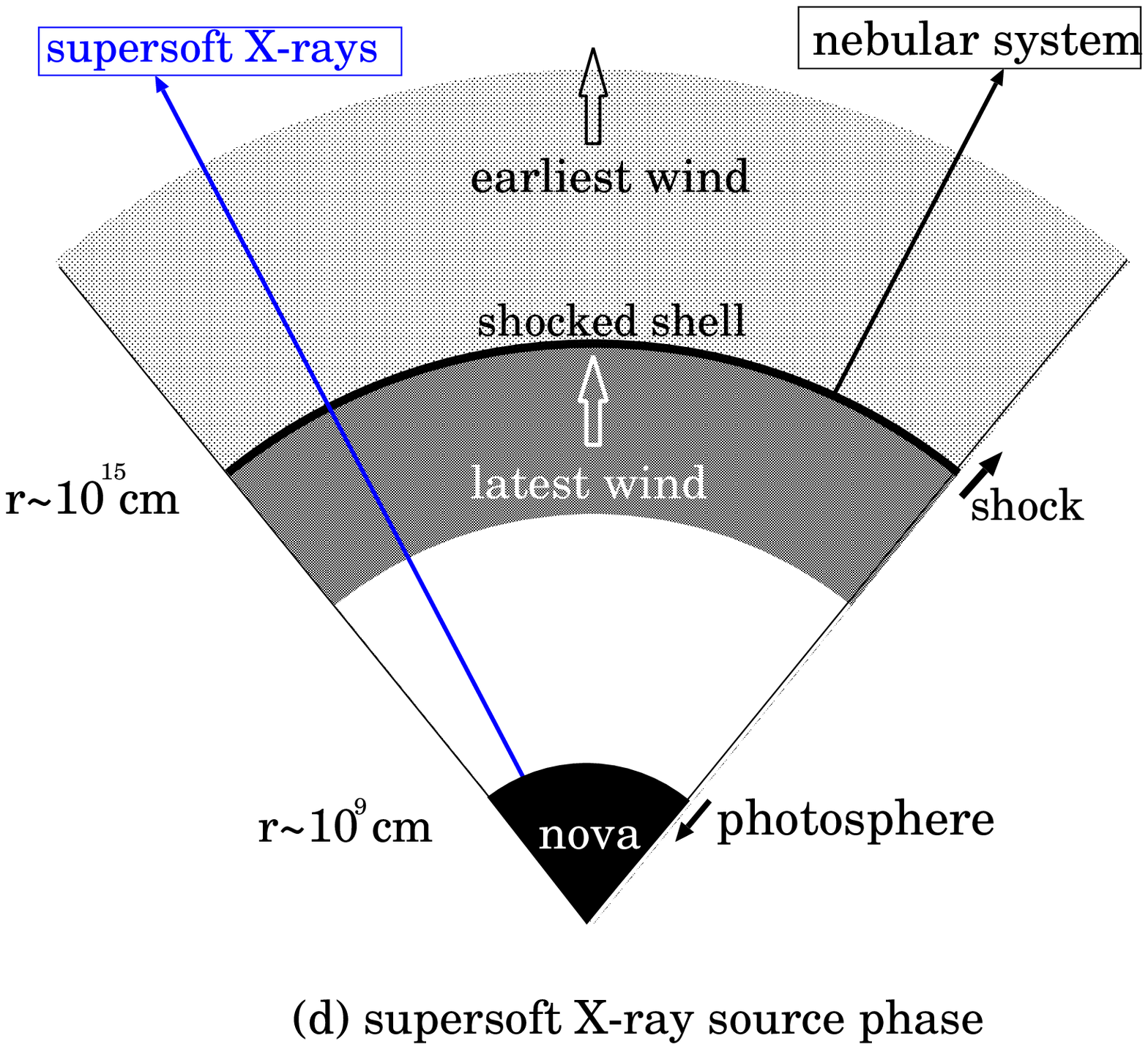}{0.45\textwidth}{}
          }
\caption{
Cartoon for our nova model.  
(a) The photospheric luminosity reaches close to the Eddington luminosity,
$L_{\rm Edd}$, with the photospheric radius being still small but
the temperature being high enough to emit supersoft X-rays.  
(b) The photosphere expands over $R_{\rm ph} \sim 0.1 ~R_\sun$ and 
optically thick winds are accelerated. 
The optically thin ejecta outside the photosphere ($r > R_{\rm ph}$) forms
the pre-maximum absorption/emission line system described in \citet{mcl42}.
(c) After optical maximum, the photosphere is receding again and
a strong shock is formed outside the photosphere \citep{hac22k}.
The shocked layer is geometrically thin, and the whole ejecta is
divided into three parts, outermost expanding gas (earliest wind),
shocked thin shell, and inner wind.  These three parts contribute to 
pre-maximum, principal, and diffuse enhanced absorption/emission line
systems in \citet{mcl42}, respectively.
The shocked shell emits thermal hard X-rays.
(d) The optically thick wind has stopped.  The photospheric temperature
becomes high enough to emit supersoft X-rays.  The latest wind still 
collides with the shocked shell.  The shocked shell contributes to
nebular absorption/emission line system.
\label{wind_shock_config}}
\end{figure*}

\section{Observational constraints}
\label{observational_constrants}

Figure \ref{optical_mass_yz_ret_x55z02o10ne03_no2} summarizes
multiwavelength light curves of optical, X-ray, and gamma-ray 
energy ranges, the data of which are all taken from \citet{kon22wa}.
In the followings, we divide the nova outburst into four stages as shown
in Figure \ref{wind_shock_config} and summarize observational properties
of YZ Ret that we should take into account in our theoretical model. 

\subsection{Pre-nova observations}
\label{pre-nova_intro}

YZ Ret was known as MGAB-V207 that showed a variation in the range
$V= 15.8$ -- 16.9 (CV mag) with two noticeable fadings down to 17.2 and 18.0
mag \citep{mur19}.  YZ Ret was classified as a nova-like VY Scl variable by
these fadings.  The mass-accretion rate onto the WD is 
${\dot M}_{\rm acc} \gtrsim ($2--3$) \times 10^{-9} ~M_\sun$ yr$^{-1}$
\citep{kat22shb}, because dwarf nova outbursts are suppressed
in a nova-like VY Scl star \citep[e.g.,][]{osa96}.

\citet{schaefer22} reported the pre-eruption orbital period of
$P_{\rm orb}= 0.1324539\pm 0.0000098$ days ($= 3.179$ hr)
based on TESS optical photometry.
The distance is estimated to be $d=2.53 \pm 0.26$ kpc
based on the Gaia early data release 3 (eDR3) \citep{bai21rf}.  
\citet{sok22ll} suggested
that YZ Ret belongs to the thick disk component of our Galaxy from 
its galactic coordinates $(\ell, b)= (265\fdg 39744, 
-46\fdg 39540)$ and the height from the galactic plane $z= -1.8$ kpc.

\subsection{X-ray flash phase}
\label{x-ray_flash_intro}
The so-called X-ray flash is a brief X-ray bright phase before optical
brightening.  Such an X-ray flash was first detected in YZ Ret with
the eROSITA instrument on board Spectrum-Roentgen-Gamma (SRG)
on UT 2020 July 7 \citep{kon22wa}.
Assuming spherical emission and the Gaia eDR3 distance,
\citet{kon22wa} obtained the X-ray luminosity (photospheric
luminosity) of $L_{\rm ph}=(2.0 \pm 1.2) \times 10^{38}$ erg s$^{-1}$
and the photospheric temperature of k$T_{\rm ph}=28.2^{+0.9}_{-2.8}$ eV, 
and photospheric radius $R_{\rm ph}= 50000\pm 18000$ km
(= $0.07~\pm 0.026~ R_\odot$) from their 36 s observation of YZ Ret.
They concluded that the duration of the X-ray flash is shorter than 8 hours.

\citet{kat22shb} presented X-ray flash light curve models
and showed that both the short duration of the X-ray flash and
the high blackbody temperature can be reproduced only in very massive
WDs ($M_{\rm WD} \gtrsim 1.3~M_\sun$) with a mass-accretion rate of
$\dot M_{\rm acc} \lesssim 5 \times 10^{-9}~M_\sun$ yr$^{-1}$.
Here, $M_{\rm WD}$ and $\dot M_{\rm acc}$ are the WD mass
and mass-accretion rate onto the WD, respectively.
We illustrate this X-ray flash phase in Figure \ref{wind_shock_config}a.

\subsection{Pre-maximum phase}
\label{pre-maximum_intro}

After the X-ray flash phase ended, the optical luminosity of the nova
increased and reached maximum.  Unfortunately, the exact time of
optical $V$ maximum cannot be well constrained.  
In what follows, we set the origin of time at $t_0 \equiv $ UT 2020 July 7,
16:47:20 $=$ JD 2,459,038.19954 $=$ MJD 59,037.69954,
which is the start time of the X-ray flash in the eROSITA observation
(see Figure \ref{optical_mass_yz_ret_x55z02o10ne03_no2}).

The nova optically brightens up from the quiescent
brightness $g\sim 16$ to $V=3.7$ at $t_0+4.06$ day \citep{mcn20}.
Here, $g$ mag are all taken from the All-Sky Automated Survey
for Supernovae (ASAS-SN) \citep{sha14pg, jay19sk}.
Because there are no optical data between $V=5.4$ at $t_0+1.08$ day
and $V=3.7$ at $t_0+4.06$ day \citep{mcn20},
the optical rise time ($\tau_{\rm peak}$)
in the $V$ band is longer than 1 days, but shorter than 4 days.
Here, $\tau_{\rm peak}$ is the duration between the X-ray flash
and the optical peak.

Such a short rise-time of $\tau_{\rm peak}$ suggests a very massive WD.
\citet{kat22shc} estimated the rise-times based on their fully
self-consistent nova outburst models.  The rise-time depends basically
on the WD mass and mass-accretion rate.  A more massive WD with a smaller 
mass accretion rate results in a shorter $\tau_{\rm peak}$.
They showed several models for two cases of $M_{\rm WD}= 1.3$ and
$1.35 ~M_\sun$, and three cases of $\dot M_{\rm acc}= 5\times 10^{-10}$,
$2\times 10^{-9}$, and $5\times 10^{-9}~M_\sun$ yr$^{-1}$.  Among them,
two models of $(M_{\rm WD}, \dot M_{\rm acc})=(1.3 ~M_\sun, 2\times 10^{-9}
~M_\sun$ yr$^{-1})$ and $(1.35 ~M_\sun, 
5\times 10^{-9} ~M_\sun$ yr$^{-1})$ 
satisfy the condition of $1 \lesssim \tau_{\rm peak} \lesssim 4$ days,
that is, $\tau_{\rm peak}=2.2$ days and 
2.9 days, respectively.
Thus, the WD mass of YZ Ret is possibly between 1.3 and 1.35 $M_\sun$
for $\dot M_{\rm acc}= (2$--$5)\times 10^{-9}~M_\sun$ yr$^{-1}$ (see
previous Sections \ref{pre-nova_intro} and \ref{x-ray_flash_intro}).

\subsection{Post-maximum phase} 
\label{post-maximumh_intro}

The $V$ light curve of YZ Ret shows a very rapid decline from $V=3.7$ 
($t_0+4.06$ day) to $V=7.2$ ($t_0+19.55$ day)
as in Figure \ref{optical_mass_yz_ret_x55z02o10ne03_no2}.
We estimated the speed class\footnote{The nova speed
class is defined by $t_3$ or $t_2$ (days of 3 or 2 mag decay from optical
maximum).  For example, 
very fast novae ($t_2 \le 10$ day), 
fast novae ($11 \le t_2 \le 25$ day), 
moderately fast nova ($26 \le t_2 \le 80$ day), 
slow novae ($81 \le t_2 \le 150$ day), 
and very slow novae ($151 \le t_2 \le 250$ day),
as defined by \citet{pay57}.}
of YZ Ret to be $t_2\approx 6$ day and $t_3\approx 13$ day, so it
belongs to the very fast nova class. 
Then, the decay slows down.  The thick green line labeled $t^{-1.75}$
indicates a power law of
\begin{eqnarray}
V &=& -2.5 \log ((t-t_0) + 10{\rm ~day})^{-1.75} -11.7 \cr
 & & + (m-M)_V + \Delta V, 
\label{nebular_emission_brightness}
\end{eqnarray}
where we adopt the power of $-1.75$ from the universal
decline law of classical novae proposed by \citet{hac06kb} and
$(m-M)_V$ is the distance modulus in the $V$ band.  For YZ Ret,
$(m-M)_V= 3.1 E(B-V) + 5 \log (d/{\rm 10~pc})= 12.1$ and $\Delta V=0$.
We adopt the Gaia eDR3 distance ($d=2.53$ kpc), and
the extinction toward YZ Ret, $E(B-V)=0.03$ after \citet{sok22ll}. 
This thick green line reasonably follows the $V$ light curve of YZ Ret
after day $t_0+18$.  We will see the reason why the same power index of 
$-1.75$ can be applied to the nebular phase in Section 
\ref{light_curve_fitting_yz_ret} and Appendix \ref{light_curve_model}.

\citet{ayd20bc} reported a high-resolution optical spectroscopy of YZ Ret
on $t_0+8.5$ day.  Their spectrum shows broad, rectangular emission lines
of Balmer, \ion{O}{1}, and \ion{Fe}{2}.  The emission lines are
accompanied by P Cygni absorption features at around $-2700$ km s$^{-1}$.
\citet{mcl21bl} also reported their dense monitoring of the optical
line profile evolution in YZ Ret.


GeV gamma-ray fluxes were detected, as shown in Figure
\ref{optical_mass_yz_ret_x55z02o10ne03_no2},
with the Fermi/Large Area Telescope (LAT) \citep{kon22wa, sok22ll}.
The flux quickly decays to non-detection level after $t_0+16.8$ day.
Note that this epoch almost coincides with the epoch
that the $V$ light curve changes from quick decay to slow decay as shown
in Figure \ref{optical_mass_yz_ret_x55z02o10ne03_no2}.

In \citet{hac22k}'s theoretical model, the gamma-ray flux emerges
after the optical peak.
The first positive detection of gamma-rays was on day $t_0+2.79$
\citep{kon22wa, sok22ll}.  The optical peak is not well
constrained ($1 \lesssim \tau_{\rm peak} \lesssim 4$ days)
unfortunately, and thus we cannot confirm the theoretical prediction.
\citet{kat22shc} discussed that the $V$ peak could
be around 2.8 days from the X-ray flash, that is,
$t_{V, \rm peak}\approx t_0+2.8$ day.
Hard X-rays were detected with NuSTAR on $t_0+10.7$ day
 \citep[filled cyan square,][]{sok20ac1, sok22ll}.


\begin{figure*}
\epsscale{0.85}
\plotone{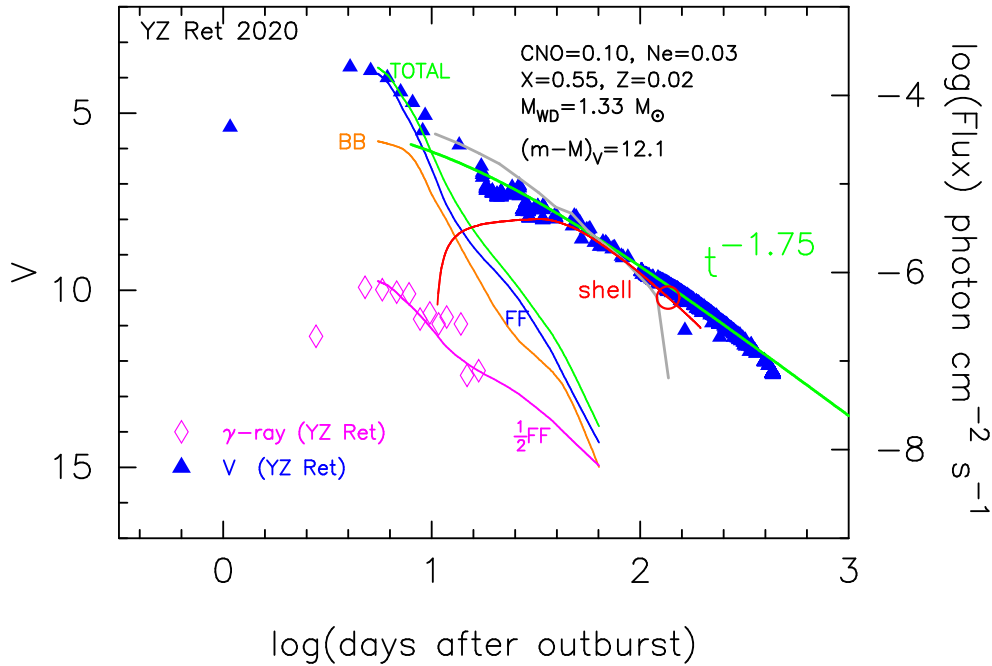}
\caption{
The optical $V$ (filled blue triangles) as well as the GeV
gamma-ray (open magenta diamonds) light curves of YZ Ret 2020.
The distance modulus in the $V$ band of $\mu_V\equiv (m-M)_V= 12.1$
is assumed.  A thick green line labeled $t^{-1.75}$ denotes
the $t^{-1.75}$ decay trend in equation (\ref{nebular_emission_brightness})
with $(m-M)_V= 12.1$ and $\Delta V= 0.0$.
We also plot our model light curves of the 1.33 $M_\sun$ WD (Ne2):
photospheric blackbody (orange line labeled BB), free-free emission
of optically thick winds (blue line labeled FF, calculated from
equation (\ref{free-free_flux_v-band})),
and the total of them (thin green line labeled TOTAL, calculated
from equation (\ref{luminosity_summation_flux_v-band})).
We further add the ${ 1 \over 2} m_{\rm ff}$ model light curve
(magenta line labeled ${1 \over 2}$FF, calculated from equation 
(\ref{gamma_ray_half_magnitude})).
The gray and red lines represent the time-stretched light curves
of \citet{kat22sha}'s fully self-consistent $1.0 ~M_\sun$ WD nova model with
the stretching factor of $\log f_{\rm s}= -0.4$.  The open red circle on the
red line indicates the epoch when optically thick winds stop.  The right edge
of the red line corresponds to the epoch where the shock disappears.
See Appendix \ref{light_curve_model} for more details of
the model light curve calcualtions.
\label{optical_mass_yz_ret_only_x55z02o10ne03_logt}}
\end{figure*}

\subsection{Supersoft X-ray source (SSS) phase}
\label{sss_phase_intro}

The X-ray flux (filled black squares) at the 0.3--2 keV band 
begins to rise drastically on $t_0+62$ day in Figure
\ref{optical_mass_yz_ret_x55z02o10ne03_no2}, 
indicating the appearance of the supersoft X-ray source (SSS) phase
\citep{sok20ac2}.  Then, it shows a flat peak between
$t_0+63$ day and $\sim t_0+90$ day. 
After day $t_0+100$, the X-ray flux begins to decay.

\citet{izz20ma} obtained the UVES spectra on $t_0 + 73$ day,
which show broad (FWZI of H$\beta \sim 4000$ km s$^{-1}$)
and strong forbidden lines emission of [\ion{O}{3}] 4959/5007 \AA, 
[\ion{O}{3}] 4363 \AA\  blended with H$\gamma$, and [\ion{N}{2}] 5755 \AA.
Balmer and He lines show a structured saddle-shaped profile suggesting
non-spherical ejecta. Their preliminary analysis shows that the observed
structure can be explained by a regular symmetric equatorial and
polar outflow at an inclination of $\sim 60\arcdeg$.
There appears to be a blue bump on all the unblended lines at 
$\sim -1250$ km s$^{-1}$, suggesting that the shell has not yet fully
settled into its 'frozen' nebular spectral line structure.
They concluded that the nova erupted on an ONe white dwarf
based on their spectra, which showed 
an overabundance of oxygen and the presence of strong
[\ion{Ne}{3}] 3342 \AA\   and [\ion{Ne}{5}] 3426 \AA\  lines.

\citet{sit20rr} reported their infrared spectroscopic observations
on $t_0+91.8$ day and $t_0+124.8$ day.  Their earlier observation showed
that YZ Ret had entered the coronal stage, displaying strong infrared
lines of [\ion{Si}{6}] and [\ion{Si}{7}], as well as [\ion{S}{8}]
and [\ion{Ca}{8}], while the later observation showed new multiple lines
of \ion{N}{5} and significantly strengthened [\ion{Si}{6}] and [\ion{Si}{7}]
lines.  Most of the emission lines showed double peaked structure with
large dips near line center, similar to what was reported
for the optical lines in \citet{izz20ma}. Full widths at half maxima
were slightly greater than 2000 km s$^{-1}$.


\begin{figure}
\gridline{\fig{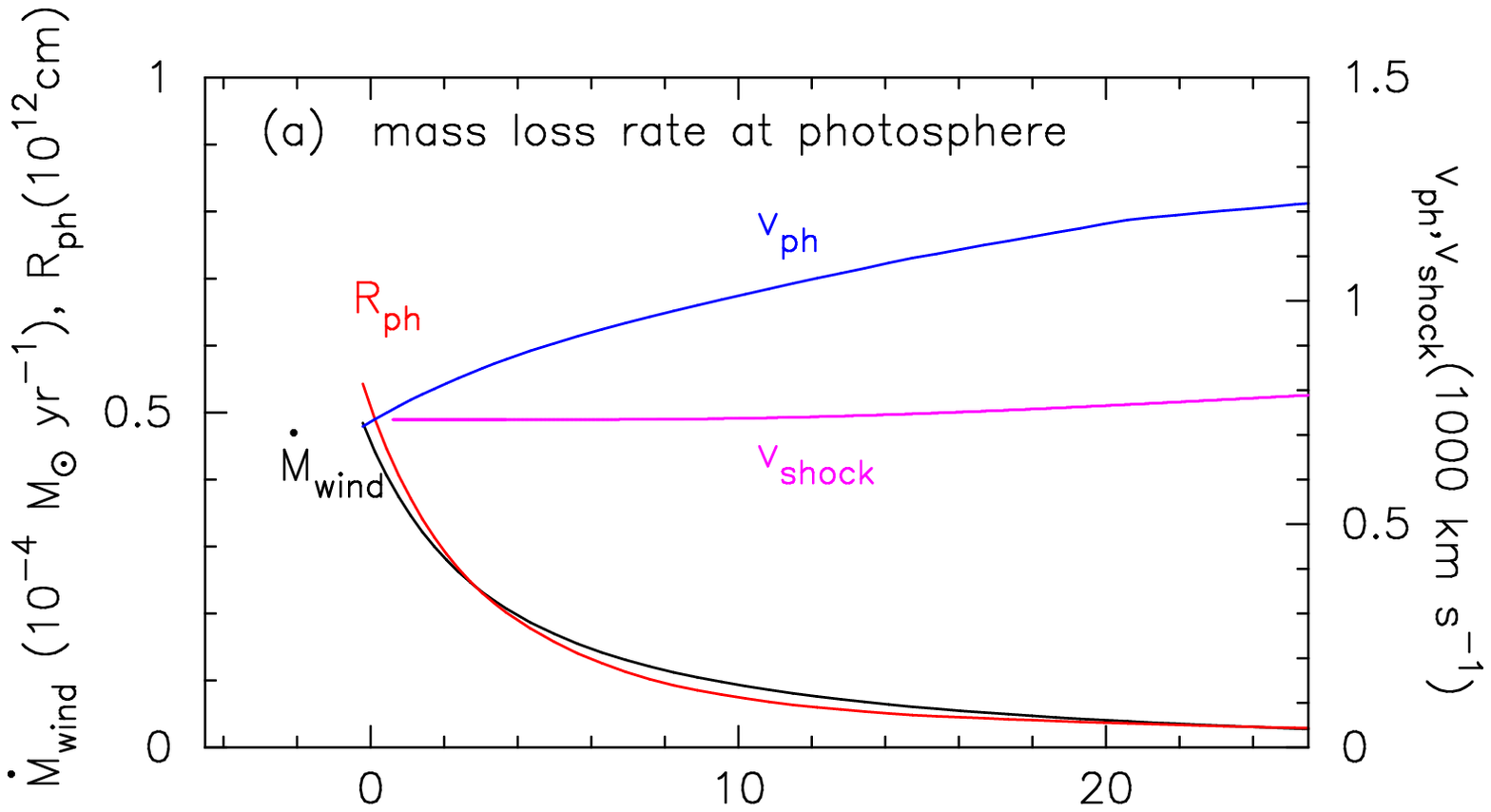}{0.47\textwidth}{}
          }
\gridline{
          \fig{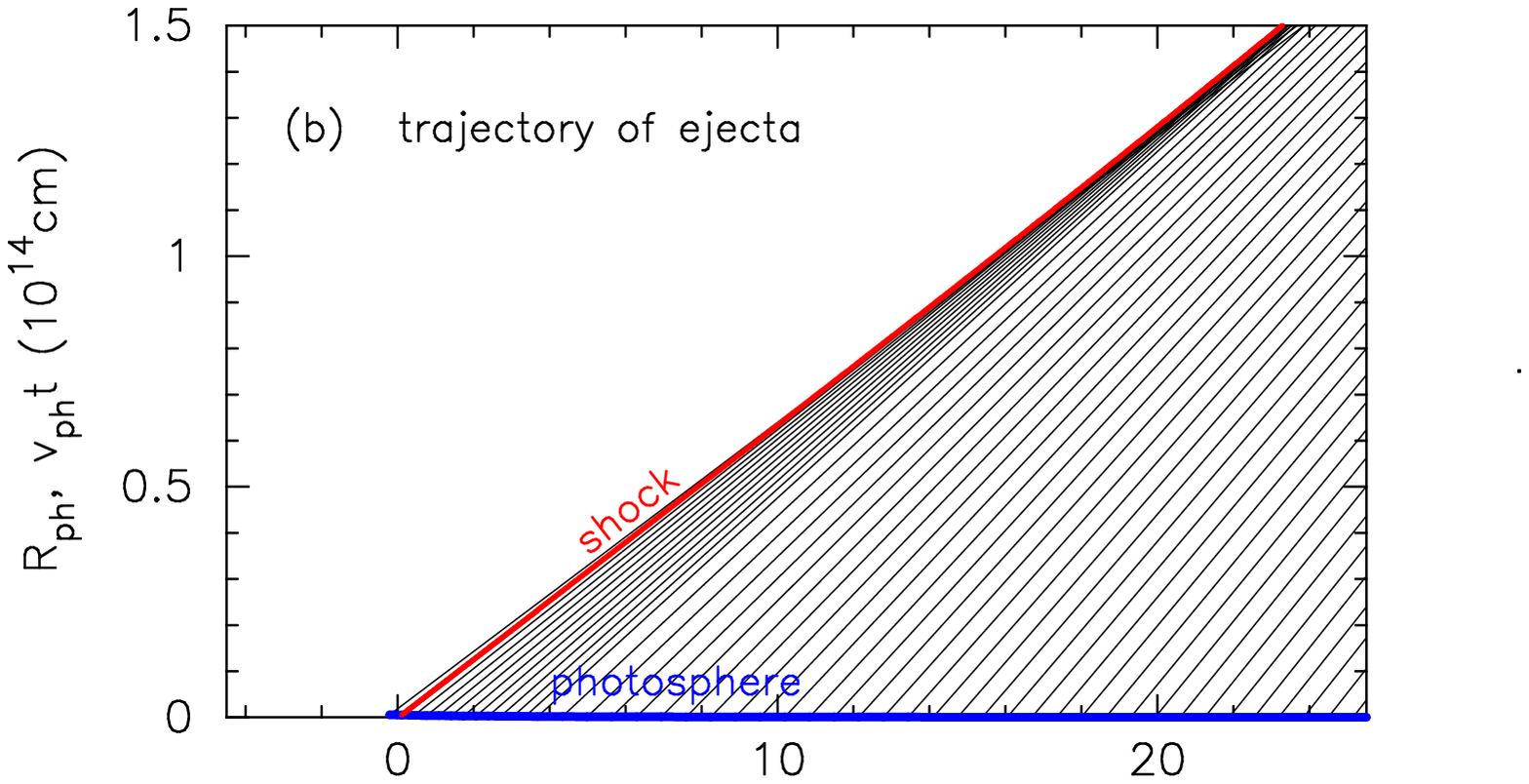}{0.46\textwidth}{}
          }
\gridline{
          \fig{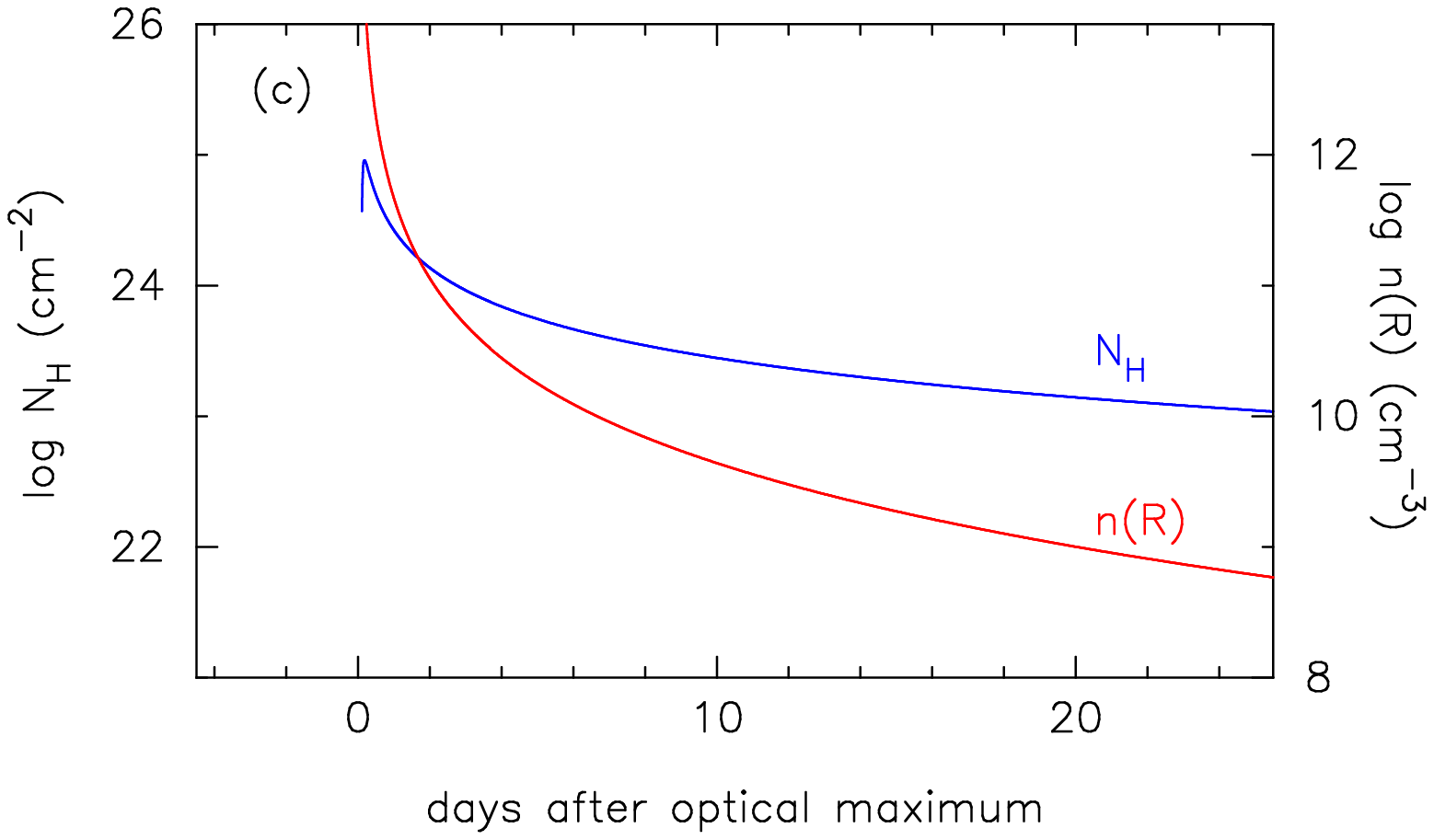}{0.47\textwidth}{}
          }
\caption{
Shock properties for our 1.33 $M_\sun$ WD (Ne2) model \citep{hac18kb}.
Only the decay phase (after the optical maximum) of steady-state envelope
model is plotted.  Each abscissa is common, days after optical maximum,
and the time $t=0$ correspond to the optical maximum.
{\bf (a)} Various photospheric properties:
the wind mass-loss rate (black line,
labeled ${\dot M}_{\rm wind}$), wind velocity (blue, $v_{\rm ph}$),
photospheric radius (red, $R_{\rm ph}$), shock velocity
(magenta, $v_{\rm shock}$).  
{\bf (b)} Trajectories (straight black lines) of winds ejected at each
time from the photosphere converge.  A strong shock (solid red line)
arises soon after the optical maximum and travels outward 
at a speed of $\sim$700--800 km s$^{-1}$.
The blue line shows the position of the photosphere.
{\bf (c)} Temporal variation in the hydrogen column density $N_{\rm H}$
(blue line) behind the shock and the number density $n(R)$
(red line) just in front of the shock ($R=R_{\rm sh}$).
\label{density_pp_interaction_yz_ret}}
\end{figure}

\section{Light curve fitting of YZ Ret}
\label{light_curve_fitting_yz_ret}

In the present paper, we study the post-maximum phase using the steady-state
wind mass-loss solutions because (1) fully self-consistent nova calculations
are numerically time-consuming and difficult and (2) the decay phase of
a nova is well described by a sequence of the steady-state wind solutions
\citep{kat22sha}.

We use steady-state wind models in \citet{hac10k}, \citet{hac16k},
and \citet{hac18kb}.  These light curve models can be applied only to
the decay phase of novae, but have been applied to the decay phases of
a number of novae and have reasonably reproduced their light curves.  
The method of model light curve fittings are described in \citet{hac15k}.
We calculated the supersoft X-ray flux assuming blackbody photosphere
for the energy range of 0.3--2 keV \citep{kat94h, hac10k}.  

\subsection{Model selection}
\label{model_selection}

Theoretical light curves of novae depend on the WD mass, chemical
composition of the envelope, and mass-accretion rate.
\citet{kat22shb, kat22shc} estimated the WD mass to be
between 1.3 and 1.35 $M_\sun$ for the mass accretion rate 
$\sim ($2--5$) \times 10^{-9}$ $M_\sun$ yr$^{-1}$
(See Sections \ref{x-ray_flash_intro} and \ref{pre-maximum_intro}). 
Therefore, we examined 1.3, 1.33, and 1.35 $M_\sun$ WDs with the
envelope chemical compositions of neon nova 2 (Ne2: $X=0.55$, $Y=0.30$,
$Z=0.02$, $X_{\rm CNO}=0.10$, $X_{\rm Ne}=0.03$) and neon nova 3 (Ne3:
$X=0.65$, $Y=0.27$, $Z=0.02$, $X_{\rm CNO}=0.03$, $X_{\rm Ne}=0.03$).
Among them, the 1.33 $M_\sun$ WD (Ne2) model well reproduces the X-ray
observation in the SSS phase (magenta lines in Figure 
\ref{optical_mass_yz_ret_x55z02o10ne03_no2}).
The X-ray flux of the 1.3 $M_\sun$ WD (Ne2) and 1.35 $M_\sun$ WD (Ne2)
models sharply increases on $t_0+82$ day and $t_0+48$ day, and these
two dates are too late and too early, respectively, 
to be consistent with the X-ray observation 
in Figure \ref{optical_mass_yz_ret_x55z02o10ne03_no2}, in which
we suppose that the X-ray turn-on time is $t_0+62$ day.
Similarly, we exclude our 1.3 $M_\sun$ WD (Ne3), 1.33 $M_\sun$ WD (Ne3),
and 1.35 $M_\sun$ WD (Ne3) models because their SSS-on day are
$t_0+112$, $t_0+82$, and $t_0+65$ day, respectively.

\subsection{Optical emission and shock formation}
\label{optical_shock_formation}

First, we calculated optical $V$ light curves in the early decay phase
($t< t_0+10$ day).  In this phase, the observed $V$ light curve is well
fitted with our model $V$ light curve (thin green line labeled TOTAL)
as shown in Figures \ref{optical_mass_yz_ret_x55z02o10ne03_no2}
and \ref{optical_mass_yz_ret_only_x55z02o10ne03_logt}.  Here,
the total $V$ luminosity $L_{V, \rm total}$ is the summation of
free-free emission $L_{V, \rm ff,wind}$ (blue line labeled FF)
and photospheric emission $L_{V, \rm ph}$ (red or orange line labeled BB). 
The calculation method is introduced in Appendix \ref{light_curve_model}.

After $t_0+10$ day, our model light curve deviates largely from the $V$
observation.  In this later phase, the shocked shell contributes to
the $V$ luminosity.

Figure \ref{density_pp_interaction_yz_ret}a depicts temporal variations
of the photospheric radius $R_{\rm ph}$, velocity $v_{\rm ph}$,
and wind mass-loss rate $\dot{M}_{\rm wind}$ as well as the shock
velocity $v_{\rm shock}$ ($=$ the velocity of the shocked shell).
Our steady-state wind model can be applied only to the post-maximum phase
\citep{hac06kb, hac15k}, so that
we plot $R_{\rm ph}$, $v_{\rm ph}$, and $\dot{M}_{\rm wind}$ only
in the decay phase of a nova, i.e., after the optical maximum.

In Figure \ref{density_pp_interaction_yz_ret}b,
we plot each locus of the wind particle, which is assumed to have a 
constant velocity after it was ejected from the photosphere. 
Each locus collides with each other and forms a shock (red line).
We also depict the hydrogen column density $N_{\rm H}$ behind the shock
front ($N_{\rm H}$ is the number density per unit area above the shock
front) in Figure \ref{density_pp_interaction_yz_ret}c.
The initial increase in $N_{\rm H}$ near $t=0$ is an accumulation
effect of the wind at the shock.
We also add the number density of gas $n(R)$ 
just in front of the shock (at the radius $R$).
Thus, we confirm that a strong shock arises soon after the optical peak
and expands at about 800 km s$^{-1}$.  Such an configuration of ejecta
is illustrated in Figure \ref{wind_shock_config}c.

The photospheric velocity $v_{\rm ph}$ decreases with time before
optical maximum and then turns to increase after that.
Figure \ref{density_pp_interaction_yz_ret}a shows $v_{\rm ph}$ of
the decay phase of our 1.33 $M_\sun$ WD.  
In the decay phase, the later ejected matter catches up
the earlier ejected gas after optical maximum (Figure 
\ref{density_pp_interaction_yz_ret}b). Thus, the collision
in the ejecta generates a shock after optical maximum. 
(See also Figure \ref{hr_mass_velocity_wind}c in Appendix 
\ref{section_nova_model}, which clearly shows that 
the ejected wind velocity ($v_{\rm ph}$)
in \citet{kat22sha}'s model is decreasing before optical maximum
but turns to increase after that.)

We estimate the flux of shell emission $L_{V, \rm shell}$ from the shocked
shell configuration (equation (\ref{luminosity_shocked_shell_ff_flux})
in Appendix \ref{light_curve_model}).
This is the contribution of free-free, bound-free, and bound-bound emissions.
In Figure \ref{optical_mass_yz_ret_only_x55z02o10ne03_logt}, we show this
flux by the red line labeled ``shell.''
The shell flux increases at $t_0+10$ day and levels off until $t_0+40$ day,
which is followed by a decline of $t^{-1.75}$, along with the thick green
line labeled $-1.75$, in the nebular phase ($t \gtrsim t_0+40$ day).
See Appendix \ref{light_curve_model} for more details of shell emission
calculation.


\subsection{Soft X-ray emission}
\label{soft_x-ray_emission}

YZ Ret entered the SSS phase on $t_0+62$ day as shown in Figure 
\ref{optical_mass_yz_ret_x55z02o10ne03_no2}.  
In our $1.33 ~M_\sun$ WD (Ne2) model, the optically-thick winds stop 
on $t_0+62$ day and the nova enters the SSS phase.
We illustrate this SSS phase in Figure \ref{wind_shock_config}d.
The photosphere of the WD envelope
shrinks to $R_{\rm ph}\lesssim 0.1 ~R_\sun$ on $\sim t_0+62$ day
and the photospheric temperature becomes high (k$T_{\rm ph} \gtrsim 25$ eV)
enough to emit soft X-rays.  
The X-ray flux (0.3--2.0 keV band) is calculated from blackbody
assumption with $T_{\rm ph}$ and $L_{\rm ph}$.
The model X-ray light curve begins to decrease on day $t_0+90$, being
consistent with the observation.


 
\citet{sok22ll} obtained X-ray spectra on $t_0+78$ day with the 
XMM-Newton/RGS, which are dominated by emission lines.
Such emission-line-dominated spectra strongly indicates a high
inclination of the binary.  The WD photosphere is occulted by the edge
of an elevated accretion disk \citep[e.g.,][]{nes13oh, sok22ll}.
It is consistent with the small X-ray fluxes during the SSS phase 
\citep[see also ][]{ori22gg}; the flux at the SSS phase is about ten times
smaller than that at the X-ray flash phase on $t_0+0$ day
(filled red square in Figure \ref{optical_mass_yz_ret_x55z02o10ne03_no2}).

\citet{izz20ma} observed YZ Ret with the UVES at ESO
on $t_0+72$ day (JD 2,459,110.5) in the mid-SSS phase.
In their high resolution spectra, Balmer and He lines show a structured
saddle-shaped profile suggesting non-spherical ejecta.
These observed structures can be explained by an 
equatorial and polar outflow at an inclination of $\sim 60\arcdeg$.
Such a high inclination angle is also
broadly consistent with the small soft X-ray flux and strong
emission-line-dominated spectra of XMM-Newton/RGS.

Therefore, we plot two X-ray light curves (thin and thick magenta lines)
in Figure \ref{optical_mass_yz_ret_x55z02o10ne03_no2}.
The thin magenta line shows the soft X-ray flux for no occultation, that is,
for the case that the absorption is the same as that for the X-ray flash
on $t_0+0$ day.
We add the soft X-ray flux (thin magenta line) at the X-ray flash of the
$1.3 ~M_\sun$ WD (on $t_0+0$ day), which is taken from \citet{kat22shb}. 
On the other hand, we plot the thick magenta line of our 1.33 $M_\sun$
WD (Ne2), the flux of which is set to follow
a flat peak of the soft X-ray between $t_0+63$ and $\sim t_0+90$ day.
This lower flux line corresponds to a twenty-seventh of 
the thin magenta line (for the case of no occulatation).
The observational fluxes are broadly located between these two lines.

\section{Emissions from a shocked shell}
\label{section_emission}

In this section, we examine how and why our $1.33 ~M_\sun$ WD model
explains shock-associated observations such as optical spectra,
hard X-rays, and GeV gamma-rays.

\subsection{Multiple velocity systems in the ejecta}
\label{multiple_velocities_ejecta}

Nova ejecta show multiple velocity systems as described by \citet{mcl42}
and \citet{mcl43}.  \citet{hac22k} interpreted \citet{mcl42}'s
pre-maximum, principal, diffuse-enhanced absorption/emission line
systems as those produced by the earliest wind before optical maximum,
shocked shell, and inner wind, as illustrated in Figure 
\ref{wind_shock_config}b and \ref{wind_shock_config}c.

YZ Ret also shows multiple velocity systems.
\citet{sok22ll} noticed two P Cygni velocity components of $-1200$ and
$-2700$ km s$^{-1}$ \citep[][and private communication]{ayd20bc} 
on JD 2,459,046.67, i.e., $t_0+8.47$ day, about 4 days post-maximum.
We regard these two velocities to be those of principal and
diffuse enhanced, $v_{\rm p}= 1200$ km s$^{-1}$ and 
$v_{\rm d}= 2700$ km s$^{-1}$, respectively, instead of 
our shock model velocities of $v_{\rm shock}= 800$ km s$^{-1}$ and
$v_{\rm wind}= 1200$ km s$^{-1}$, which tend to be smaller than
the observation.
Here, $v_{\rm p}$ is the velocity of the principal system and
$v_{\rm d}$ that of the diffuse enhanced system.


\citet{sok22ll} estimated the ejecta mass in YZ Ret to be between
$M_{\rm ej}= 2\times 10^{-6} ~M_\sun$
(for assumed $v_{\rm shell}= 1200$ km s$^{-1}$ and 
$N_{\rm H}= 7.3 \times 10^{22}$ cm$^{-2}$) and
$M_{\rm ej}= 2\times 10^{-4} ~M_\sun$
(for assumed $v_{\rm shell}= 2700$ km s$^{-1}$ and 
$N_{\rm H}= 131.3 \times 10^{22}$ cm$^{-2}$),
where $v_{\rm shell}= v_{\rm shock}= v_{\rm p}$
is the velocity of the shocked shell.
Our $1.33 ~M_\sun$ WD model suggests $M_{\rm ej}= 4\times 10^{-6} ~M_\sun$.
We also obtain $N_{\rm H}\sim 30 \times 10^{22}$ cm$^{-2}$ 
(for $v_{\rm shell}= 800$ km s$^{-1}$) 
from Figure \ref{density_pp_interaction_yz_ret}c.
Our estimates are broadly consistent with the above values.

\subsection{Duration of shock}
\label{shock_duration_hard_x-ray}

In our nova model, a shock arises just after the optical peak (see, e.g.,
Figures \ref{wind_shock_config}, \ref{density_pp_interaction_yz_ret}, 
and Appendix Figure \ref{hr_mass_velocity_wind}).
The shock disappears when the wind stops.  More accurately,
the shock ends when the last wind reaches the shock front.
The shock front is far outside the photosphere, so it takes
$t_{\rm ret}$ days from the date of wind stopping at the photosphere
to the time of shock termination. 
In the last day of the shock, the shock front is at $R_{\rm sh}(t)$.
The winds collide with themselves forming the shock.
The wind velocity is $v_{\rm wind}(t)= v_{\rm ph}(t-t_{\rm ret})$
at $R_{\rm sh}(t)$, because it was ejected
from the photosphere $t_{\rm ret}$ days before the collision. 
This retarded (look back) time is
$t_{\rm ret}= (R_{\rm sh}(t) - R_{\rm ph}(t-t_{\rm ret}))
/v_{\rm ph}(t-t_{\rm ret})$, where $R_{\rm ph}(t-t_{\rm ret})$
and $v_{\rm ph}(t-t_{\rm ret})$ are the photospheric radius and velocity
at $t-t_{\rm ret}$.  Assuming that the last wind was emitted at the wind
stopping time $t_{\rm ws}$, the shock duration of 
$\tau_{\rm shock}\equiv t_{\rm ws}+ t_{\rm ret}$ is calculated from
\begin{equation}
(t_{\rm ws}+ t_{\rm ret}) v_{\rm sh}(t_{\rm ws}+ t_{\rm ret}) 
=  t_{\rm ret} v_{\rm ph}(t_{\rm ws}), 
\end{equation}
where $v_{\rm sh}$ is the velocity of the shocked shell at
$t_{\rm ws}+ t_{\rm ret}$, $v_{\rm ph}$ the velocity 
at the photosphere at $t_{\rm ws}$.  
This simply means that the shock front moves with the speed of $v_{\rm sh}$
from the optical maximum and the last wind catches up the shock with
the speed of $v_{\rm ph}$.

Our model predicts the shock duration by 
substituting $v_{\rm sh}\approx v_{\rm p}=1200$ km s$^{-1}$
(principal system), $v_{\rm ph}\approx v_{\rm d}=2700$ km s$^{-1}$
(diffuse enhanced system), and
$t_{\rm ws}=58$ day (the wind duration just after the shock arises)
into 
\begin{equation}
\tau_{\rm shock}= {{t_{\rm ws}} \over
{\left( 1- {{v_{\rm p}} \over {v_{\rm d}}}\right)}}.
\label{duration_of_shock}
\end{equation}
The shock duration is $\tau_{\rm shock}= 58/0.4444= 104$ day.
Therefore, we expect hard X-ray emission until about $t_0+110$ day in YZ Ret.

\citet{izz20ma} pointed out that a blue bump appears at $\sim -1250$
km s$^{-1}$ on all the unblended lines in their $t_0+72$ day observation. 
Thus the shell has not yet fully settled into its 'frozen' nebular structure.
On the other hand, \citet{gal20m} reported that the nova shell has reached
its 'frozen' structure on $t_0+110$ day.
Thus, the freezing occurred between $t_0+72$ day and $t_0+110$ day. 
This epoch is consistent with our result that the shock disappeared 
before $t_0+110$ day and then the shell entered the free expansion phase
(no acceleration by shock). 

\citet{mcl21bl} presented rapidly evolving spectroscopic sequences
of YZ Ret, especially of the H$\alpha$ complex, from $t_0+8.76$ day 
to $t_0+175$ day.  They found that the evolution trend of
the H$\alpha$ velocity features changed on $t_0+100$ day.
This epoch is broadly coincident with the end of shock in our model
as mentioned above.


\subsection{Rectangular line profiles: jets vs. shocked shell}
\label{shell_versus_jets}

Many novae showed a rectangular shape of emission lines in the nebular
phase.  \citet{bea31} analyzed flat-top emission lines in the two classical
novae, V476 Cyg 1920 and V603 Aql 1918, and concluded that such a 
rectangular profile arises from an optically-thin, spherical thin shell.
The line width of a rectangle shape corresponds to $2 v_{\rm shell}$.

In the present paper, we assume that the optically-thick winds and
shocked shell are spherically symmetric as illustrated in Figure 
\ref{wind_shock_config}c and \ref{wind_shock_config}d.
YZ Ret also showed a rectangular-like shape of emission lines of
Balmer, \ion{O}{1}, and \ion{Fe}{2} \citep{ayd20bc, mcl21bl}.
This supports that an optically-thin, spherical shell is formed
in the nova ejecta.  Note that many photo-images of nova shells show
almost spherically- or ellipsoidally-distributed blobs like in GK Per 1901 
\citep[e.g.,][]{sha12zd}.  \citet{sla95od} showed that the axis ratio of
a nova shell is related to the nova speed class
such that slower novae appear to have
more elongated shells.  Thus, spherically symmetric assumption is
broadly consistent with observational features of the very fast nova class
of YZ Ret ($t_2\approx 6$ day).

On the other hand, \citet{mcl21bl} proposed a 'jets' model.
They decomposed an emission line with a dip at the line center into two
Gaussians; blue-shifted and red-shifted components, which correspond to
near-coming and far-receding jets, respectively.  If it is the case, 
many lines are originated from the jets.  This simply means that a large
part of the nova envelope is ejected in the jets.  As a result,
many novae should show a bi-polar geometry in the ejecta, which is
against observations.  See also, e.g., \citet{hut72} for relations
between line profiles with dips and non-spherical shells.

We have a possible reason why a line profile has a dip at the line center
even for a spherical shell.
If the geometrically-thin shell has small optical thickness (small but
finite line optical depth $\tau$), the rectangular line profile becomes
a ``saw-toothed'' shape with a dip at the line center 
\citep[e.g.,][]{wag83bb, ber87m}.

\citet{sok22ll} also argued against \citet{mcl21bl}'s jets model for YZ Ret.
Their points are two-fold: (1) the line-dominated SSS phase in YZ Ret
strongly suggests a high inclination angle of the binary.  The observed
velocities of $\sim 1000$ km s$^{-1}$ is too high to be compatible with
the projected velocities of jets.  (2) If blobs in the jet collide with
each other and emits X-rays, its timescale is determined by the blob size,
that should be short.  This is against the observed long timescale.
Therefore, the emitting region is relatively large.

\subsection{Hard X-ray emission}
\label{section_hard_x-ray}

The temperature just behind the shock is estimated to be
\begin{eqnarray}
kT_{\rm sh}& \sim & {3 \over 16} \mu m_p 
\left( v_{\rm wind} - v_{\rm shock} \right)^2 \cr
& \approx & 1.0 {\rm ~keV~} 
\left( {{v_{\rm wind} - v_{\rm shock}} \over  
{1000 {\rm ~km~s}^{-1}}} \right)^2,
\label{shock_kev_energy}
\end{eqnarray}
where $k$ is the Boltzmann constant,
$T_{\rm sh}$ is the temperature just after the shock
\citep[see, e.g.,][]{met14hv}, 
$\mu$ is the mean molecular weight ($\mu =0.5$ for hydrogen plasma),
and $m_p$ is the proton mass.
Substituting $v_{\rm shock}= v_{\rm p}=1200$ km s$^{-1}$ and 
$v_{\rm wind}= v_{\rm d}=2700$ km s$^{-1}$,
we obtain the post-shock temperature 
$k T_{\rm sh}\sim 2.3$ keV.

Mechanical energy of the wind is converted to thermal energy
by the reverse shock \citep{met14hv} as
\begin{eqnarray}
L_{\rm sh}& \sim & {{9}\over {32}} {\dot M}_{\rm wind} 
{{( v_{\rm wind} - v_{\rm shock} )^3} \over {v_{\rm wind}}} \cr
&=& 1.8\times 10^{37}{\rm ~erg~s}^{-1}
\left( {{{\dot M}_{\rm wind}} \over 
{10^{-4} ~M_\sun {\rm ~yr}^{-1}}} \right) \cr
 &  & \times
\left( {{{v_{\rm wind} - v_{\rm shock}} \over {1000{\rm ~km~s}^{-1}}}}
\right)^3
\left( {{{1000{\rm ~km~s}^{-1}} \over {v_{\rm wind}}} }\right). 
\label{shocked_energy_generation}
\end{eqnarray}
Substituting $\dot{M}_{\rm wind}= 0.5 \times 10^{-4} ~M_\sun$ yr$^{-1}$
from Figure \ref{density_pp_interaction_yz_ret}a,
we obtain the post-shock energy of
$L_{\rm sh} \sim 1.1\times 10^{37}$ erg s$^{-1}$.

\citet{sok22ll} analyzed the NuSTAR observation of YZ Ret on $t_0+10.7$ day.
They fitted a single temperature thermal plasma model with the observed
spectrum (3.5--78.5 keV) and obtained $kT=6.5\pm 1.5$ keV and 
$N_{\rm H}=20^{23}$--$10^{24}$ cm$^{-2}$, depending on abundances.
Our $1.33 ~M_\sun$ WD (Ne2) model gives
$kT_{\rm sh}= 2.3$ keV and $N_{\rm H}\sim 2 \times 10^{23}$ cm$^{-2}$
on the same day, which are broadly consistent with the results 
by \citet{sok22ll}.

In the later nebular and SSS phases, the velocity and mass of
shocked shell do not change so much.
The column density of hydrogen is estimated from
$M_{\rm shell}= 4 \pi R_{\rm sh}^2 \rho h_{\rm shell}$,
where
$\rho$ is the density
in the shocked shell, and $h_{\rm shell}$ the thickness of the shocked shell.
If we take an averaged velocity of shell $v_{\rm sh}=
v_{\rm shell}= v_{\rm shock}$,
the shock radius is calculated from $R_{\rm sh}(t)= v_{\rm shock}\times t$.
This reads
\begin{eqnarray}
N_{\rm H} & = & {{X \over m_p} {{ M_{\rm shell} }
\over {4 \pi R^2_{\rm sh}}}} \cr
 & \approx & 4.8\times 10^{22} {\rm ~cm}^{-2}
\left({X \over {0.5}}\right)
\left( {{M_{\rm shell}} \over {10^{-5} M_\sun}} \right)
\left( {{R_{\rm sh}} \over {10^{14} {\rm ~cm}}} \right)^{-2}
\cr
 & \approx & 6.4 \times 10^{20} {\rm ~cm}^{-2}
\left({X \over {0.5}}\right)
\left( {{M_{\rm shell}} \over {10^{-5} M_\sun}} \right) \cr
& & \times
\left( {{v_{\rm shell}} \over {1000 {\rm ~km~s}^{-1}}} \right)^{-2}
\left( {{t} \over {100~{\rm day}}} \right)^{-2}.
\label{column_density_hydrogen_time}
\end{eqnarray}
This gives $N_{\rm H}\approx 3\times 10^{20}$ cm$^{-2}$ for
$M_{\rm shell}= 0.5\times 10^{-5} ~M_\sun$, 
$v_{\rm shell}=1200$ km s$^{-1}$, and $t=90$ day, which is the end
of the supersoft X-ray source phase.
  

\citet{pei20og} and \citet{ori22gg} 
analyzed the NICER observation (0.3--2 keV) of YZ Ret
on $t_0+83$ and $t_0+84$ day and obtained $T_{\rm ph}=$500,000--550,000 K
and $N_{\rm H}= 3\times 10^{20}$ cm$^{-2}$,
which is consistent with our estimate mentioned above.

\subsection{GeV gamma-ray emission}
\label{gev_gamma_ray_emission}


Recently gamma-rays were often detected during nova outbursts
\citep[e.g.,][]{ack14}.  GeV gamma-rays were also observed in the
YZ Ret outburst with the Fermi/Large Area Telescope (LAT) \citep{sok22ll}. 
Such gamma-rays are considered to originate from
strong shocks \citep[see, e.g.,][for a recent review]{cho21ms}.
The gamma-ray flux peaked on JD 2,459,042.99 ($t_0+ 4.79$ day) \citep{kon22wa},
slightly later than the brightest $V$ observation
($V=3.7$ at $t_0+4.06$ day),
as already introduced in Section \ref{pre-maximum_intro}.
This epoch of appearance is consistent with our shock model.

The gamma-ray fluxes (open magenta diamonds) in Figure
\ref{optical_mass_yz_ret_only_x55z02o10ne03_logt}
broadly follows the trend of optical $V$ decay. 
In our model, the optical flux is dominated by free-free emission
and is given by $L_{V, \rm ff}\propto (\dot{M}_{\rm wind}/v_{\rm ph})^2
\propto f(t)$ (equation (\ref{free-free_flux_definition})) 
while the gamma-ray flux is given by 
$L_{\gamma, \rm sh} \propto (\dot{M}_{\rm wind}/v_{\rm wind})\propto 
[f(t)]^{1/2}$
(equation (\ref{shocked_energy_generation})), where $f(t)$ is a function
of time $t$.  Thus, the dependencies of $L_{V, \rm ff}$ and
$L_{\gamma, \rm sh}$ on $(\dot{M}/v)^n$ is similar, 
but the power of $n$ is different by a factor of two ($n=2$ vs $n=1$).
Therefore, the light curves in magnitudes are written as
\begin{eqnarray}
m_{\gamma, \rm sh}(t) &=& -2.5 \log L_{\gamma, \rm sh} + {\rm ~const.} \cr
&=& -2.5 \log [f(t)]^{1/2} + {\rm ~const.} \cr
&=& {1\over 2}\left(-2.5 \log f(t) + {\rm ~const.}\right) \cr
&=& {1\over 2}(-2.5 \log L_{V, \rm ff} + {\rm ~const.}) \cr
&=& {1\over 2} m_{V,\rm ff}(t) + {\rm ~const}., 
\label{gamma_ray_half_magnitude}
\end{eqnarray}
so the decay slope is slow by a factor of 2.
Here, the proportionality constant is included in the term ``const.''
We plot the $m_{\gamma, \rm sh}$ (cyan line labeled ${1 \over 2}$FF)
in Figures \ref{optical_mass_yz_ret_x55z02o10ne03_no2} and
\ref{optical_mass_yz_ret_only_x55z02o10ne03_logt}.  This model slope
represents well the decay trend of GeV gamma-ray fluxes.

\citet{sok22ll} obtained the flux of GeV gamma-ray (0.1--300 GeV) to be 
$L_{\gamma}= 3\times 10^{35}$ erg s$^{-1}$ and the optical luminosity
to be $L_{\rm opt}= 7\times 10^{38}$ erg s$^{-1}$ on $t_0+4.1$ day.
The gamma to optical ratio is $L_{\gamma}/L_{\rm opt} = 4\times 10^{-4}$.
In our model, the shock energy generation is 
$L_{\rm sh}=1.1\times 10^{37}$ erg s$^{-1}$.
The ratio of $L_{\gamma}/L_{\rm sh}= 0.027$,
about 3\% conversion rate, is consistent with
\begin{equation}
L_{\gamma}= \epsilon_{\rm nth} \epsilon_{\gamma} L_{\rm sh}
\lesssim 0.03 ~L_{\rm sh},
\end{equation}
where $\epsilon_{\rm nth}\lesssim 0.1$ is the fraction of the shocked
thermal energy to accelerate nonthermal particles and
$\epsilon_{\gamma}\lesssim 0.1$ the fraction
of this energy radiated in the Fermi/LAT band \citep[typically 
$\epsilon_{\rm nth} \epsilon_{\gamma} < 0.03$,][]{met15fv}.



\begin{figure}
\epsscale{1.15}
\plotone{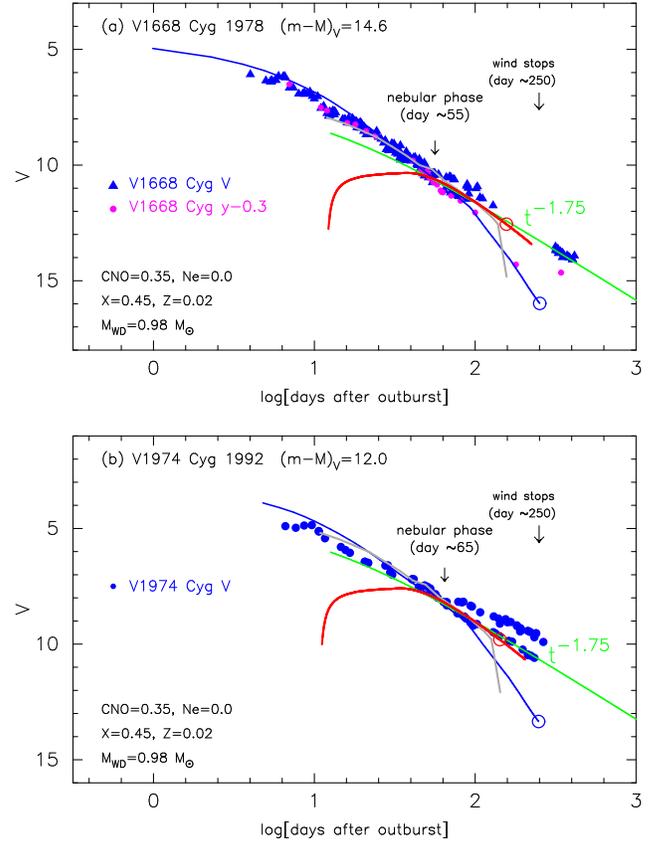}
\caption{
The $V$ (and $y$) light curves of V1668 Cyg (a) and V1974 Cyg (b).
The $V$ and $y$ data are the same as those in \citet{hac16kb}.
The blue lines denote the total $V$ luminosity of our models.
The open blue circle at the right edge of each blue line
corresponds to the epoch when optically thick winds stop.
The gray and red lines indicate the $V$ luminosities of wind and
shell emission, respectively, of \citet{kat22sha}'s fully self-consistent
nova model, as summarized in Appendicies 
\ref{section_nova_model} and \ref{light_curve_model}. 
The open red circle on the red line indicates
the epoch when optically thick winds stop.  The right edge of the red line
corresponds to the epoch when the shock disappears.
{\bf (a)} V1668 Cyg.  We adopt the time-stretching factor of 
$\log f_{\rm s}= -0.34$ for the fully self-consistent nova model
(gray and red lines).  The green line labeled $t^{-1.75}$ shows
equation (\ref{nebular_emission_brightness}) with $(m-M)_V= 14.6$
and $\Delta V= -0.2$ 
{\bf (b)} V1974 Cyg.  We adopt $\log f_{\rm s}= -0.38$.
The green line labeled $t^{-1.75}$ denotes equation
(\ref{nebular_emission_brightness}) with $(m-M)_V= 12.0$
and $\Delta V= -0.2$. 
\label{v1668_cyg_v1974_cyg_x45z02c15o20_calib_log}}
\end{figure}

\section{Extension of the universal decline law to the nebular phase}
\label{t175_law_nebular_phase}

\citet{hac06kb} theoretically proposed a $t^{-1.75}$ decay of free-free
emission luminosity.  This decline law explains many classical novae
light curves.  Therefore, it has been named the universal decline law.
However, \citet{hac06kb}'s model light curve of $L_{V, \rm ff}$ (blue line) 
begins to deviate from the $V$ observation after the nova entered
the nebular phase, as in Figures
\ref{v1668_cyg_v1974_cyg_x45z02c15o20_calib_log},
\ref{v1500_cyg_v959_mon_x45z02c15o20_calib_log},
\ref{v496_sct_v2659_cyg_x55z02c10o10_calib_log}, and Appendix Figure
\ref{optical_m10_abs_lv_vul_mag}.  This clearly suggests that
emissions from the shocked shell contribute significantly to
the $V$ flux of the nova after the nebular phase starts.
We suppose that equation (\ref{nebular_emission_brightness}) represents
a temporal variation of the $V$ flux from such a shell emission.  

It should be noted here that the universal decline law indicates
$L_{V, \rm ff}\propto (t-t_0)^{-1.75}$.  This curve is slightly different
from that of equation (\ref{nebular_emission_brightness}), because 
we derived equation (\ref{nebular_emission_brightness}) from the direct
fit with the YZ Ret $V$ light curve for $t\gtrsim t_0+10$ day.
Large differences from the $(t-t_0)^{-1.75}$ universal decline law come
from the term $(t-t_0+ 10 {\rm ~day})^{-1.75}$ in equation
(\ref{nebular_emission_brightness}).  However, the difference between
the two laws becomes small for $t-t_0 \gg 10 {\rm ~day}$.
 
In this section, we explain how this trend of $L_V \propto t^{-1.75}$
can be applied to the nebular phase of the other novae.
The flux of shell emission is calculated based on \citet{kat22sha}'s
fully self-consistent nova model.  We apply this shell luminosity to
each nova with the time-stretching method \citep{hac18kb}:
two nova light curves of different speed class
overlap each other, if the timescale of one of them is squeezed by
a factor of $f_{\rm s}$, i.e., as $t/f_{\rm s}$, and its absolute $V$
brightnesses is shifted to be $M_V-2.5\log f_{\rm s}$.
In other words, the two nova light curves overlap each other 
in the $(t/f_{\rm s})$-$(M_V-2.5\log f_{\rm s})$ plane.
(See appendix \ref{light_curve_model} for more detail.)
We pick up seven well observed novae among
what we have already studied.  
LV Vul is examined as an example in appendix \ref{light_curve_model}.
We show the other six novae as follows:
The red line in each figure denotes the shell luminosity, whereas
the blue line indicates our model flux of  $L_{V, \rm total}=
L_{V, \rm ff} + L_{V, \rm ph}$ calculated from optically thick winds.
The red line is for \citet{kat22sha}'s 1.0 $M_\sun$ WD model, 
but we overlap it with each nova light curve with the time-stretching 
method.

\subsection{V1668 Cyg 1978}
\label{v1668_cyg_1978}


Figure \ref{v1668_cyg_v1974_cyg_x45z02c15o20_calib_log}a
shows the wide $V$ band magnitudes (filled blue triangles) and
intermediate $y$ band magnitudes (filled magenta circles) of V1668 Cyg.
We adopt a $0.98 ~M_\sun$ WD (CO3) model light curve \citep{hac16k}.
In the figure, our model light curve (blue line) well
follows the $V$ and $y$ light curves from day $\sim 4$ to day $\sim 100$,
while the $V$ and $y$ light curves depart from each other on day $\sim 100$.

The model light curve (blue line) follows the
$y$ light curve rather than the $V$ light curve.
While the wide $V$ band includes strong emission lines such as [\ion{O}{3}],
the free-free emission model light curve does not include
them (strong emission lines).  On the other hand, the intermediate $y$ band
is designed to avoid strong emission lines such as [\ion{O}{3}] 
\citep[see, e.g., Figure 1 of ][]{mun13dc} and follows continuum flux
\citep{loc76m}.

The $V$ light curve itself seems to split slightly on day $\sim 55$.  
This difference comes from the sensitivity of each filter system.
In the nebular phase, the strong nebular [\ion{O}{3}] 4958.9
and 5006.9 \AA\  emission lines are located just at the blue edge of 
the $V$ bandpass \citep[see, e.g., Figure 1 of ][]{mun13dc}
and a small difference in the sensitivities of the two
filters make a substantial difference in their $V$ magnitudes. 
This kind of split in the $V$ magnitude at the different observatories
can be seen on day $\sim 65$ in V1974 Cyg (see Figure 
\ref{v1668_cyg_v1974_cyg_x45z02c15o20_calib_log}b).
Thus, the splitting in $V$ magnitude indicates the beginning of
the nebular phase.

In Figure \ref{v1668_cyg_v1974_cyg_x45z02c15o20_calib_log}a, our model
light curve (blue line) crosses the red line on day $\sim 55$.  
After that, the shell emission dominates the total $V$ flux of the nova.  
We overplot the $t^{-1.75}$ decline law (green line)
of equation (\ref{nebular_emission_brightness}) with
$(m-M)_V= 14.6$ and $\Delta V= -0.2$.
This line represents well the $V$ light curve in the nebular phase
of V1668 Cyg (after day $\sim 55$).   

\subsection{V1974 Cyg 1992}
\label{v1974_cyg_1992}
  
Figure \ref{v1668_cyg_v1974_cyg_x45z02c15o20_calib_log}b
show the $V$ band magnitudes (filled blue circles) of V1974 Cyg.
We adopt a $0.98 ~M_\sun$ WD (CO3) model light curve \citep{hac16k}.
In the figures, our model light curve (blue line) well follows 
the $V$ light curves from day $\sim 6$ to day $\sim 100$.
The $V$ light curve splits into two branches on day $\sim 65$,
where the nebular phase started \citep[e.g.,][]{hac16kb}.

In Figure \ref{v1668_cyg_v1974_cyg_x45z02c15o20_calib_log}b,
The blue line crosses the red line on day $\sim 65$, suggesting
that the shell luminosity dominates the $V$ flux of the nova.
This is almost the same day that the nebular phase started. 

We also overplot the $t^{-1.75}$ decline law (green line)
of equation (\ref{nebular_emission_brightness}) with
$(m-M)_V= 12.0$ and $\Delta V= -0.2$.
This green line follows both the red line and the lower branch of the $V$
light curve in the nebular phase of V1974 Cyg (after day $\sim 65$).   


\begin{figure}
\epsscale{1.15}
\plotone{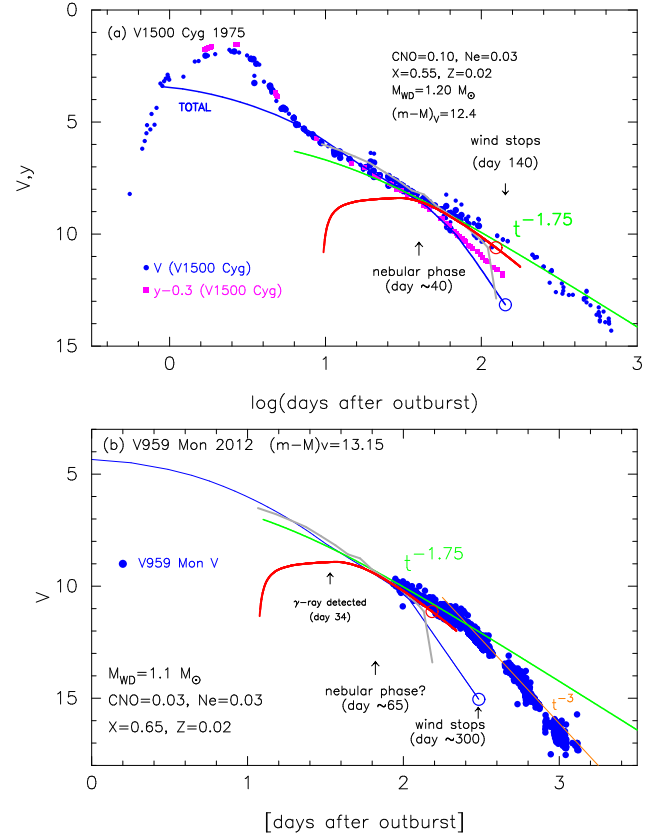}
\caption{
Same as Figure \ref{v1668_cyg_v1974_cyg_x45z02c15o20_calib_log},
but for V1500 Cyg and V959 Mon.
{\bf (a)} V1500 Cyg.  The $V$ and $y$ data are the same as those in
\citet{hac16kb}.  We adopt the time-stretching factor of 
$\log f_{\rm s}= -0.44$ for the fully self-consistent nova model
(gray and red lines).  The green line shows
equation (\ref{nebular_emission_brightness})
with $(m-M)_V= 12.4$ and $\Delta V= +0.3$.   
{\bf (b)} V959 Mon.  The $V$ data are taken from \citet{hac18k}.
The time-stretching factor is $\log f_{\rm s}= -0.35$.
The green line shows equation (\ref{nebular_emission_brightness})
with $(m-M)_V= 13.15$ and $\Delta V= -0.35$.
The decline law of $\propto t^{-3}$ (orange line)
denotes emission from a freely-expanding shell after the shock disappeared.
\label{v1500_cyg_v959_mon_x45z02c15o20_calib_log}}
\end{figure}

\subsection{V1500 Cyg 1975}
\label{v1500_cyg_1975}

Figure \ref{v1500_cyg_v959_mon_x45z02c15o20_calib_log}a shows the wide
$V$ band magnitudes (filled blue circles) and intermediate $y$ band
magnitudes (open magenta circles) of V1500 Cyg.  The early 1--4 day
broadband spectra can be fitted well with a black body \citep[e.g.,][]{gal76}.
Then, the optical and IR spectra changed from blackbody to free-free
emission on day 4--5 (after the outburst) as discussed by \citet{gal76}
and \citet{enn77}.   In the figures, our model light
curve (blue line) well follows the $V$ and $y$ light curves from
day $\sim 5$ to day $\sim 40$.

We adopt a $1.2 ~M_\sun$ WD (Ne2) model \citep{hac14k}.
The $V$ and $y$ light curves depart from each other on day $\sim 40$,
where the nebular phase starts \citep[e.g.,][]{hac16kb}.

\citet{hac16k} obtained the start of the nebular phase to be $V=8.5$, where
the $y$ light curve (open magenta circles) separates from the $V$ light
curve (filled blue circles).  Our model
light curve (blue line) follows the $y$ light curve rather than
the $V$ light curve as shown in Figure 
\ref{v1500_cyg_v959_mon_x45z02c15o20_calib_log}a.

As for the other sources, we add the shell emission light curve which
exceeds the blue line on day $\sim 40$.
This is almost the same day that the nebular phase started.  Therefore,
in the nebular phase, the shell emission dominates the $V$ flux of the nova.
We also add the $t^{-1.75}$ decline law (green line)
of equation (\ref{nebular_emission_brightness}) with
$(m-M)_V= 12.4$ and $\Delta V= +0.3$.
This green line follows both the red line and the $V$ light curve
in the nebular phase of V1500 Cyg (after day $\sim 60$).   

\subsection{V959 Mon 2012}
\label{v959_mon_2012}
  
Figure \ref{v1500_cyg_v959_mon_x45z02c15o20_calib_log}b shows the $V$ band
magnitudes (filled blue circles) of V959 Mon.  This nova was detected
in GeV gamma-rays with the Fermi/LAT  \citep{ack14}
before the optical discovery by S. Fujikawa
on UT 2012 August 9.8 (JD~2,456,149.3) at mag 9.4 \citep{fuj12}.
Due to solar conjunction, the nova already entered
the nebular decline phase when it was discovered.
The optical peak was possibly substantially (more than 50 days)
before the discovery \citep[e.g.,][]{mun13dc}.

We adopt a $1.1 ~M_\sun$ WD (Ne3) model light curve \citep{hac18k}.
In the figure, our day 0 (outburst day)
is set to be JD 2,456,066.5 after \citet{hac18k},
34 days before the first gamma-ray detection at JD 2,456,100.5
(= UT 2012 June 22).  The model light curve begins to deviate from
the $V$ light curves from day $\sim 100$, where the nebular phase had
already started \citep{mun13dc}.
Comparing the $y$ and $V$ magnitude developments,
\citet{mun13dc} concluded that a bifurcation between $y$ and $V$ light
curves took place at the start of supersoft X-ray source phase, $\sim 200$
days after the outburst.  After that, strong emission lines such as
[\ion{O}{3}] 4958.9 and 5006.9 \AA\  begins to contribute to the $V$ band
magnitudes \citep[see, e.g., Figure 1 of ][]{mun13dc}.


As for the other emission, 
the shell emission line crosses the blue line on day $\sim 65$, suggesting
that the shell emission luminosity dominates the $V$ flux of the nova.
The nebular phase possibly started near above day $\sim 65$. 
We also overplot the $t^{-1.75}$ decline law (green line)
of equation (\ref{nebular_emission_brightness}) with
$(m-M)_V= 13.15$ and $\Delta V= -0.35$.
This green line follow the $V$ light curve in the nebular phase
until day $\sim 260$.  

After that, the $V$ light curve declines
along $L_V \propto t^{-3}$ line (orange line labeled $t^{-3}$),
where $L_V$ is the $V$ band flux.
This rapid decline can be understood from a new condition
that the thickness of the shell increases with time in equation 
(\ref{luminosity_shocked_shell_ff_flux}), i.e., 
$h = h_0 \times (t/260 ~{\rm day})$ for $t > 260$ day.
On day $\sim 300$, optically-thick winds stop and we expect that the 
shock and the resultant compression at the shocked shell becomes weak,
which is not able to keep the thickness $h$ nearly constant.


\begin{figure}
\epsscale{1.15}
\plotone{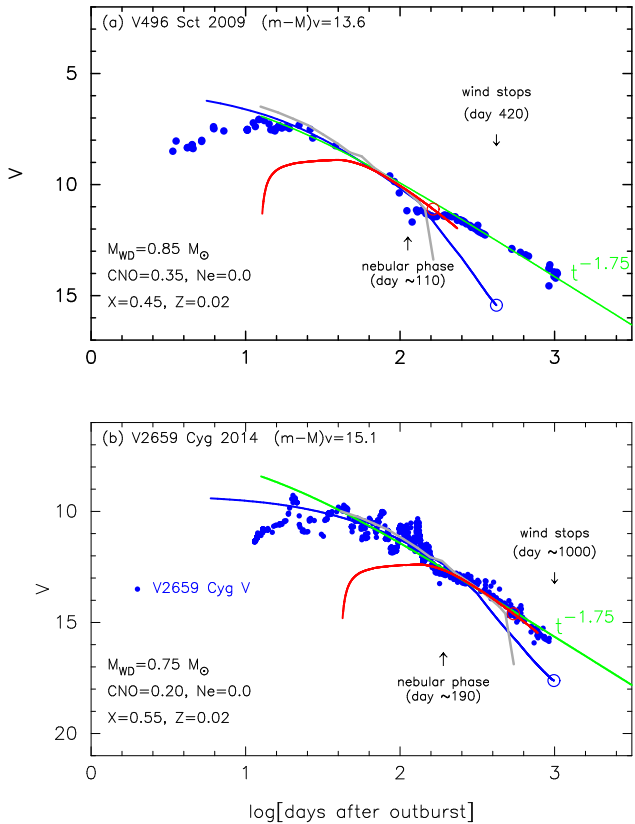}
\caption{
Same as Figure \ref{v1668_cyg_v1974_cyg_x45z02c15o20_calib_log},
but for V496 Sct and V2659 Cyg.  The $V$ data of V496 Sct are
taken from AAVSO and \citet{raj12ab}, while those of V2659 Cyg are 
the same as those in \citet{hac21k}.
(a) V496 Sct.  We adopt the time-stretching factor of 
$\log f_{\rm s}= -0.32$.  The green line shows
equation (\ref{nebular_emission_brightness}) 
with $(m-M)_V= 13.6$ and $\Delta V= -0.9$.
(b) V2659 Cyg.  The time-stretching factor is $\log f_{\rm s}= +0.2$.
The green line represents equation 
(\ref{nebular_emission_brightness}) with $(m-M)_V= 15.1$
and $\Delta V= -0.9$.
\label{v496_sct_v2659_cyg_x55z02c10o10_calib_log}}
\end{figure}

\subsection{V496 Sct 2009}
\label{v496_sct_2006}

Figure \ref{v496_sct_v2659_cyg_x55z02c10o10_calib_log}a
shows the $V$ band magnitudes (filled blue circles) of V496 Sct.
The $V$ data are taken from \citet{raj12ab} and the archive of
the American Association of Variable Star Observers (AAVSO).
We adopt a $0.85 ~M_\sun$ WD (CO3) model and the distance modulus in
the $V$ band of $(m-M)_V=13.6$ from \citet{hac21k}.
Our model light curve (blue line) broadly follows the $V$ light curve 
from day $\sim 20$ to day $\sim 100$, although the $V$ observation has
a small dip around day $\sim 110$.  Dust grows in the shocked shell
and absorbs optical fluxes both from free-free emission and 
shell emission.  This absorption by a dust shell is the so-called ``dust 
blackout.''  Here, we set the outburst day (day 0)
as JD 2,455,142.0 (= UT 2009 November 6.5), 12.2 days before 
the optical maximum at $V=7.07$ \citep{raj12ab}.

\citet{raj12ab} obtained a spectrum in the nebular phase 114 days
after optical maximum, that is, day 126 in our Figure
\ref{v496_sct_v2659_cyg_x55z02c10o10_calib_log}a.
This date is close to the epoch of a shallow dust blackout.
Thus, the nebular phase started around day $\sim 110$.
Our total flux of BB+FF (blue line) also depart from the $V$
(filled blue circles) light curve around the same epoch.

The blue line crosses the red line on day $\sim 110$, suggesting
that the shell luminosity dominates the $V$ flux of the nova.
We also overplot the $t^{-1.75}$ decline law (green line)
of equation (\ref{nebular_emission_brightness}) with
$(m-M)_V= 13.6$ and $\Delta V= -0.9$.  The green line follows 
the red line at least from day $\sim 80$ to day $\sim 160$,
and the $V$ light curve from day $\sim 200$ to day $\sim 1000$.

\subsection{V2659 Cyg 2014}
\label{v2659_cyg_2014}
  
Figure \ref{v496_sct_v2659_cyg_x55z02c10o10_calib_log}b
shows the $V$ band magnitudes (filled blue circles) of V2659 Cyg.
We adopt a $0.75 ~M_\sun$ WD (CO4) \citep{hac21k}.
The model light curve (blue line) reasonably
follows the $V$ light curve from day $\sim 15$ to day $\sim 300$,
although the $V$ light curve is not smoothly declining
but sometimes shows spikes and flares.
Here, we set the outburst day as JD 2,456,737.0, 18.5 days before
the $V$ maximum at $V=9.4$ mag \citep[e.g.,][]{tar16}.
\citet{tar16} mentioned, based on her optical spectra 146 days 
after the $V$ maximum, that the nova was in an early phase of the
nebular stage.  Her next spectra were obtained on 166 and 192 days
after the $V$ maximum, which clearly show nebular emission lines such
as [\ion{O}{3}].  Therefore, the nova entered the nebular phase 
$\sim 160-170$ days after the optical maximum.

The blue line crosses the red line on day $\sim 190$, suggesting
that the shell luminosity dominates the $V$ flux of the nova.
The nebular phase possibly started near day $\sim 190$, which is
consistent with the start day proposed
by \citet{tar16}, i.e., $170+18.5 \approx 190$ days after the outburst.
We also overplot the $t^{-1.75}$ decline law 
(green line) of equation (\ref{nebular_emission_brightness}) 
with $(m-M)_V= 15.1$ and $\Delta V= -0.9$.  This green line follows
the $V$ light curve in the nebular phase of V2659 Cyg.





\section{Conclusions}
\label{conclusions}

We proposed a multiwavelength light curve model for the classical nova 
YZ Ret 2020, especially for the decay phase.
Our main results are summarized as follows:\\

\noindent
{\bf 1.} 
Our model $V$ light curve of free-free emission for a $1.33 ~M_\sun$ WD (Ne2)
well reproduces the early decay phase until 
$t_0+10$ day.
\\

\noindent
{\bf 2.}
After $t_0+10$ day, we modeled a $V$ light curve supposing that
the optical emission mainly comes from the shocked shell.  We found that
the flux follows approximately the decay trend of $\propto t^{-1.75}$.
This is the same trend as the universal decline law proposed by 
\citet{hac06kb} for free-free emission.  Thus, the decline trend of
$L_V \propto t^{-1.75}$ can be applied from the early decline phase
(dominated by free-free emission of winds) through the nebular phase 
(dominated by shocked shell emission) until the shock disappears. 
\\

\noindent
{\bf 3.} 
The same 1.33 $M_\sun$ WD model reproduces the X-ray light curve of YZ Ret,
from $t_0+62$ day to $\sim t_0+100$ day, i.e., the supersoft X-ray source (SSS)
phase.  The peak flux of the SSS phase is about a tenth of the flux at
the X-ray flash.  Because we directly observed the WD photosphere
at the X-ray flash, the low flux during the SSS phase can be explained
by an occultation of the WD surface.
\\

\noindent
{\bf 4.}
A nova ejecta is divided by the shock into three parts,
outermost expanding gas (earliest wind before maximum),
shocked shell, and inner fast wind.
These three are responsible for the pre-maximum, principal,
and diffuse enhanced absorption/emission line systems, respectively.
We interpret that the shock velocity $v_{\rm shock}$
corresponds to the velocity $v_{\rm p}$ of the principal system and
the inner wind velocity $v_{\rm wind}$ to the velocity $v_{\rm d}$ of the
diffuse enhanced system.  The shock temperature is calculated to be
$k T_{\rm sh} \sim 2.3$ keV from equation (\ref{shock_kev_energy}),
assuming $v_{\rm p}= 1200$ km s$^{-1}$ and $v_{\rm d}= 2700$ km s$^{-1}$.
This temperature and the hydrogen column density behind the shock
on $t_0+10.7$ day are roughly consistent with
the estimates by \citet{sok22ll}.
\\

\noindent
{\bf 5.} 
Our shocked energy post-maximum is $L_{\rm sh}\sim 1.1\times 10^{37}$
erg s$^{-1}$, calculated from equation (\ref{shocked_energy_generation}).
The observed GeV gamma-ray energy is $L_{\gamma}\sim 3\times 10^{35}$
erg s$^{-1}$ \citep{sok22ll}.
The ratio of $L_{\gamma}/L_{\rm sh}\sim 0.027$ satisfies the theoretical
request \citep[$L_{\gamma}/L_{\rm sh} \lesssim 0.03$, ][]{met15fv}.
\\

\noindent
{\bf 6.} 
The shocked energy generation rate depends roughly on 
$\propto {\dot M}_{\rm wind}/v_{\rm wind}$ and the free-free
emission flux depends differently on 
$\propto {(\dot M}_{\rm wind}/v_{\rm wind})^2$.
Therefore, the decay trend of gamma-ray is roughly half as fast as
that of optical,
i.e., $m_{\gamma, \rm sh}= {1 \over 2}m_{V, \rm ff} +$const. in magnitude.
This relation broadly reproduces the decay of GeV gamma-ray flux 
observed with Fermi/LAT.
\\

\noindent
{\bf 7.} 
We estimate the duration of the shock alive based on our shock model
from equation (\ref{duration_of_shock}),
that is, the shock continues from $t_0+3$ day to $t_0+110$ day.
The end of the shock
indicates that the shocked shell begins to expand freely without
significant acceleration.  The event causes the freezing of emission 
line features.  This epoch of $t_0+110$ day is consistent with
the spectroscopic observations by \citet{izz20ma}, \citet{gal20m},
and \citet{mcl21bl}. 
\\

\noindent
{\bf 8.}
The above shocked-shell-emission model can be applied to the nebular
phase of other novae.
We show that the decline trend of $L_V \propto t^{-1.75}$ is common in
the nebular phases of V1668 Cyg, V1974 Cyg, V1500 Cyg, V959 Mon, 
V496 Sct, V2659 Cyg, and LV Vul.
\\

\begin{acknowledgments}
We are grateful to the anonymous referee
for detailed and useful comments regarding how to improve the manuscript.
We also thank 
the American Association of Variable Star Observers (AAVSO) and
the Variable Star Observers League of Japan (VSOLJ)
for the archival data of various novae that we analyzed.
\end{acknowledgments}

\appendix
\section{One cycle of a nova outburst}
\label{section_nova_model}

In the present paper we adopt the 1.33 $M_\sun$ WD (Ne2) model for YZ Ret. 
This model is based on the steady-state approximation, that can be applied
only to the decay phase (i.e., after optical maximum) of a nova outburst.  
The decay phase is only a part of one cycle of a nova outburst. 
This appendix gives a supplemental information for a nova outburst evolution.

\subsection{HR diagram} \label{appendix_HR}

Figure \ref{hr_mass_velocity_wind}a shows the HR diagram 
for one cycle of a nova outburst \citep{kat22sha}. 
This model is calculated with the fully 
consistent evolution method, in which the
internal structure, from the white dwarf (WD) center to the 
photosphere, is consistently connected with the outer wind solution. 
Winds are accelerated by radiation pressure gradients, 
deep inside the photosphere, the so-called optically thick winds. 
This model is for a $1.0~M_\sun$ WD with
the mass accretion rate to the WD before the outburst of
${\dot M}_{\rm acc}= 5\times 10^{-9}$ $M_\sun$ yr$^{-1}$, 
different from our YZ Ret model, but 
this is only the fully self-consistent model of 
a classical nova outburst. 
The WD mass and mass-accretion rate are typical for classical novae, 
so we take this model as a representative of a nova outburst 
in the following explanation. 

In the present work, we analyzed the YZ Ret observation for 
four different stages as in Figure \ref{wind_shock_config}. 
These stages correspond to four colored regions in the 
HR diagram (Figure \ref{hr_mass_velocity_wind}a), that is, the
X-ray flash (red), rising phase (cyan blue), decay phase (blue), and 
SSS phase (orange). 

Timescale of each phase and physical values 
depend on the WD mass and mass-accretion rate. 
In the $1.0 ~M_\sun$ WD model, the temperature 
at point C is $3.8\times 10^5$ K, 
which is insufficient to emit strong X-ray flux. 
\citet{kat22sha} calculated the X-ray and UV light curves and 
showed most of the energy is emitted in the UV band, and the 
X-ray flux is very weak. Thus, no X-ray flash is expected 
in less massive WDs than $1.0 ~M_\sun$.  In more massive WDs 
($> 1.3~M_\sun$), on the other hand, their one cycle loops are
located on the upper-left side of the $1.0 ~M_\sun$ WD. 
The temperature $T_{\rm ph}$ at point C increases up to 
600,000--1,000,000 K and a very bright X-ray flash is expected. 
The duration of the flash is very short in massive WDs 
\citep{kat22shb, kat22shc}.

The maximum temperature in the X-ray flash (point C) is lower than 
that in the SSS phase (Figure \ref{hr_mass_velocity_wind}a). 
This property is common for all the WD masses and mass accretion rates. 
Thus, the peak luminosity of the X-ray flash 
should be lower than that in the SSS phase. 
Nevertheless, in YZ Ret, the observed X-ray flux in the SSS phase 
is about ten times smaller than in the X-ray flash detected 
with the SRG/eROSITA \citep{kon22wa}, as in Figure 
\ref{optical_mass_yz_ret_x55z02o10ne03_no2}.
Thus, the reason for this lower flux could be an occultation 
of the WD photosphere/surface by an elevated disk.
This is consistent with the emission-line-dominated
spectra in the SSS phase \citep{sok22ll}.

\subsection{Optically Thick Winds}

There is no indication of wind mass-loss in the X-ray flash phase 
of YZ Ret \citep{kon22wa} and the spectrum is well represented 
by a blackbody spectrum of $T=3.3 \times 10^5$ K. 
This is consistent with the theoretical calculation for a 1.0 $M_\sun$ WD
\citep{kat22sha} in which optically thick winds are not accelerated
in the very early phase but begin to blow when the photospheric
temperature decreases to point E in Figure \ref{hr_mass_velocity_wind}a. 

In the very early phase of shell flash (from point B to C), 
convection widely occurs in the envelope that carries nuclear energy 
efficiently to the surface and the envelope does not expand. 
The envelope begins to expand when the envelope becomes radiative 
from the surface region. In the radiative region, opacity is an
important factor for wind acceleration. 
The radiative opacity has a large peak corresponding to the iron ionization 
at $T \sim 2 \times 10^5$ K \citep{igl96r}.
In the case of our $1.0 ~M_\sun$ WD, optically thick winds start to blow
at point E ($\log T$ (K)= 5.319).  As the envelope expands and 
the photospheric temperature decreases, the density in the wind 
acceleration region increases with time. 
Thus, the wind mass loss rate increases, too.

Stage G in Figure \ref{hr_mass_velocity_wind}a is the maximum expansion
of the photosphere.  The photospheric temperature reaches minimum.
The wind mass-loss rate reaches maximum.  The wind velocity
at the photosphere decreases and reaches minimum at the same epoch.
Their temporal variations are plotted in Figure \ref{hr_mass_velocity_wind}b.


The velocities in Figure \ref{hr_mass_velocity_wind}b 
are smaller than typical absorption line
velocities of fast novae (1000--2000 km s$^{-1}$) 
but consistent with those of slow and very slow novae 
(200--800 km s$^{-1}$) as in Figure 11 of \citet{ayd20ci}.
The velocity of wind $v_{\rm ph}$ at the photosphere
decreases toward the maximum expansion of the photosphere and then increases.
This trend in the photospheric velocity $v_{\rm ph}$ is commonly observed
in many novae \citep[e.g.,][]{ayd20ci}.  

\subsection{Formation of a strong shock outside the photosphere}
\label{internal_shock_formation}

Strong shocks have been suggested from recent detection
of gamma-rays and hard X-rays in nova outbursts \citep[e.g.,][]{ack14}. 
Such high energy photons are assumed to originate 
from internal shocks in nova ejecta 
\citep[e.g.,][]{met14hv, met15fv, mar18dj, cho21ms}. However, 
no theoretical calculations of thermonuclear runaway show a shock wave 
formation, i.e., no shock arises inside the envelope including 
the nuclear burning region up to the photosphere. It is 
mainly because the timescale of nuclear energy generation ($\sim 100$ s)
is much longer than the hydrodynamic timescale ($\sim 0.4$ s) in the
hydrogen-burning zone \citep[see][for more detail]{kat22sha, hac22k}.

\begin{figure*}
\gridline{\fig{f8a.eps}{0.425\textwidth}{(a)}
          \fig{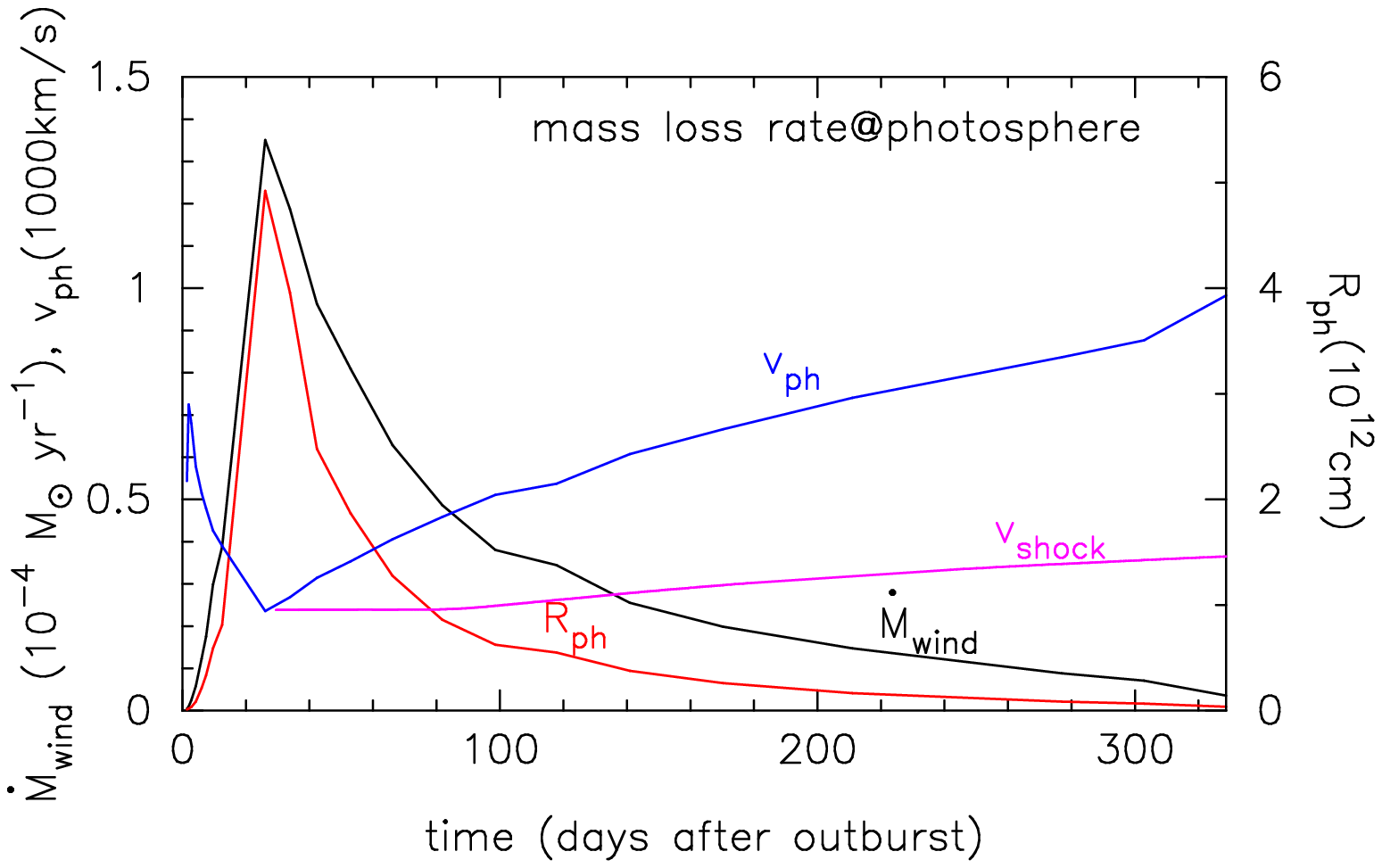}{0.5\textwidth}{(b)}
          }
\gridline{
          \fig{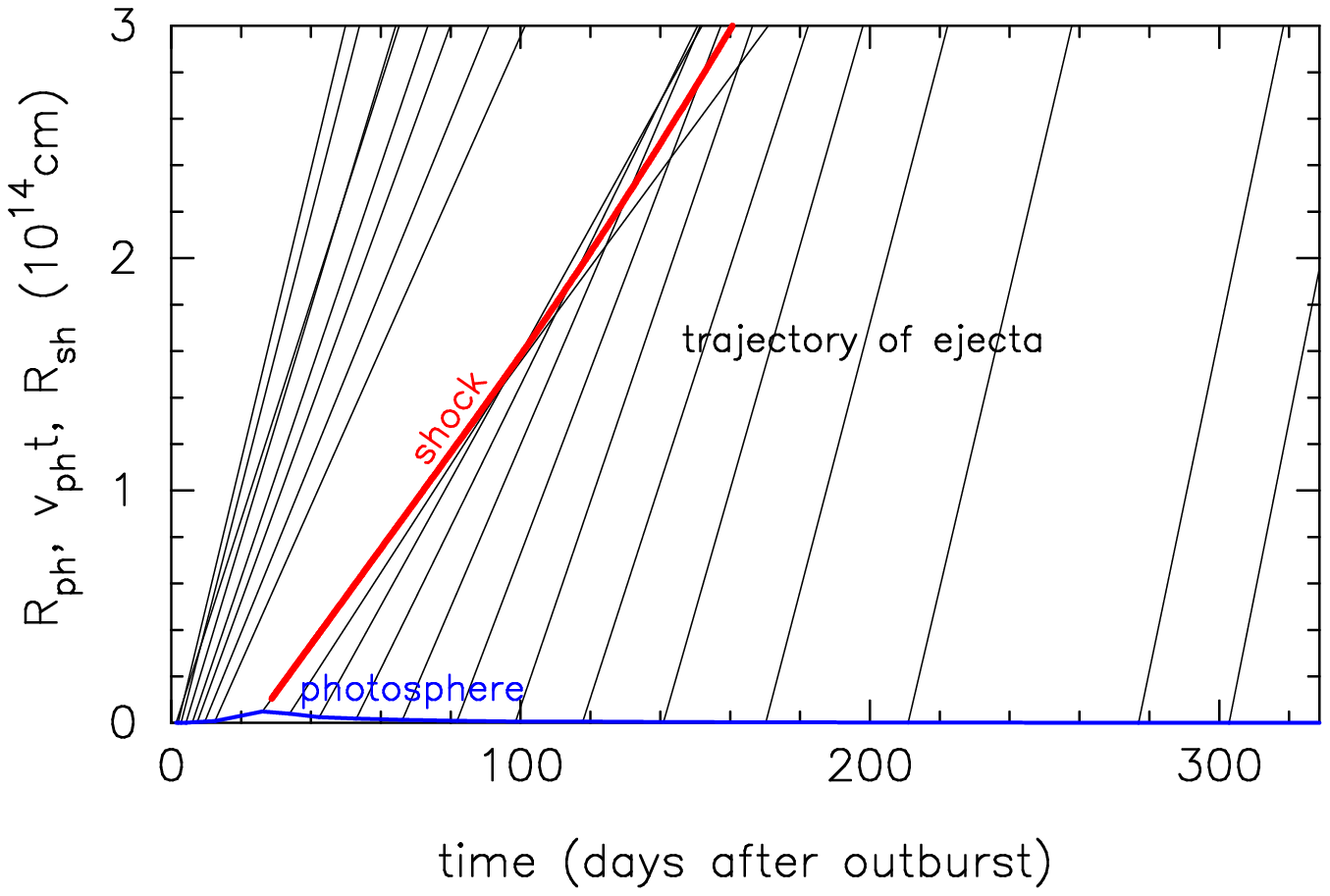}{0.425\textwidth}{(c)}
          \fig{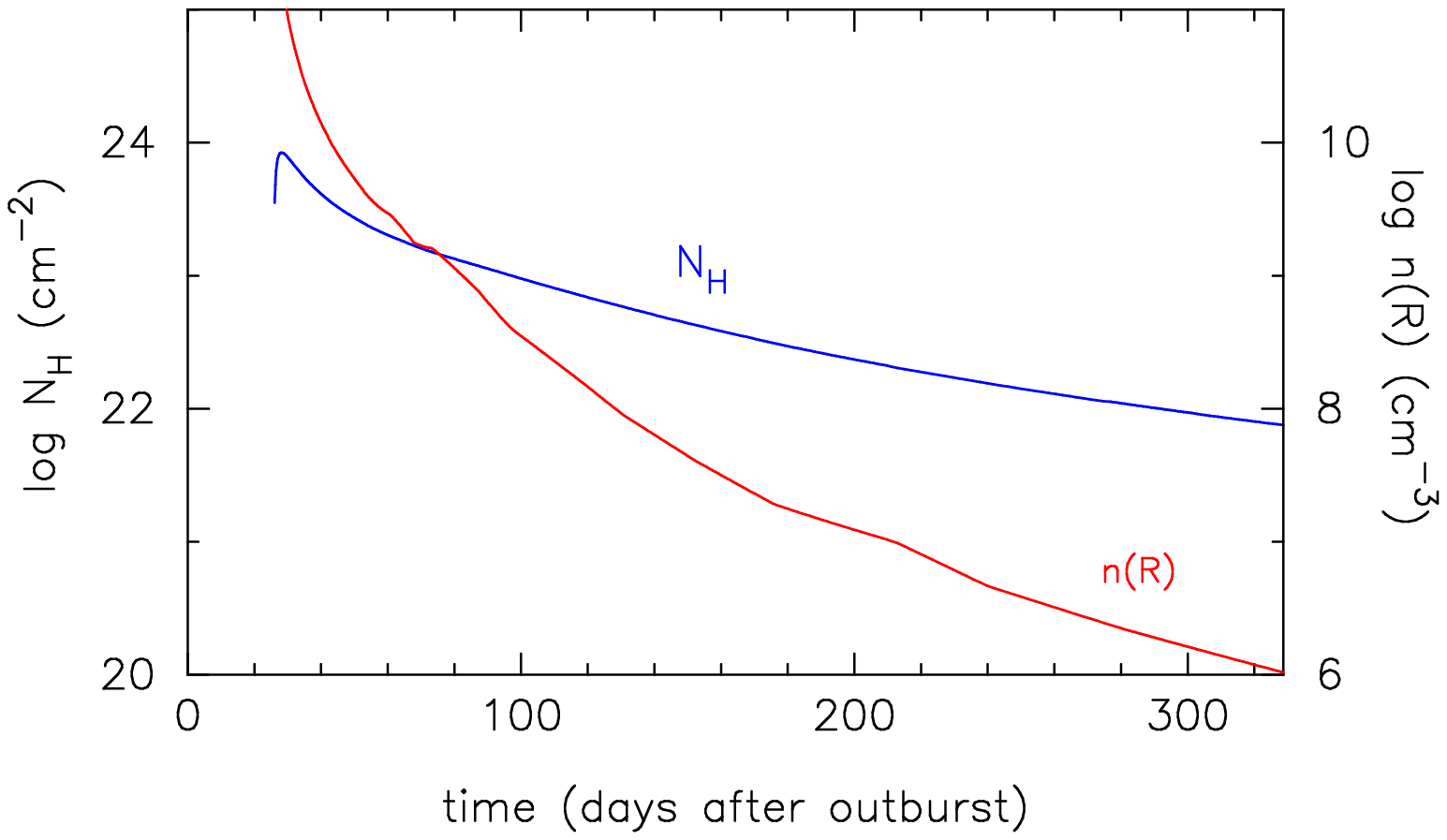}{0.5\textwidth}{(d)}
          }
\caption{
A fully self-consistent nova model on a $1.0 ~M_\sun$ WD with a mass
accretion rate of $5\times 10^{-9} ~M_\sun$ yr$^{-1}$, which is taken from
\citet{kat22sha}.  A shock model corresponding to this nova model
is taken from \citet{hac22k}.
{\bf (a)} The HR diagram. 
A: Quiescent phase before a shell flash.
B: The epoch when the nuclear luminosity reaches maximum ($t=0$).
C: The photospheric temperature takes the maximum value in the rising phase.
E: The wind mass loss begins.
F: The epoch at $\log T_{\rm ph}$ (K)=5.0.
G: The epoch when the photospheric radius reaches maximum.  At the same time,
the wind mass-loss rate attains the peak value.
H: The epoch at $\log T_{\rm ph}$ (K)=4.3.
I: The wind mass loss stops.
J: The supersoft X-ray luminosity decreases to one tenth of the maximum value.
{\bf (b)} Temporal variations of various photospheric properties:
the wind mass-loss rate (black line,
labeled ${\dot M}_{\rm wind}$), wind velocity (blue, $v_{\rm ph}$),
photospheric radius (red, $R_{\rm ph}$), shock velocity
(magenta, $v_{\rm shock}$).  
{\bf (c)} Propagation of the shock front (thick red line) and
trajectories (thin black lines) of the winds.
We add the position of the photosphere (thick blue line).
{\bf (d)} Temporal variation in the hydrogen column density $N_{\rm H}$
(blue line) behind the shock and the number density $n(R)$
(red line) just in front of the shock ($R=R_{\rm sh}$).
\label{hr_mass_velocity_wind}}
\end{figure*}

\citet{hac22k} solved this problem based on \citet{kat22sha}'s
self-consistent nova model.  They showed that a strong shock arises
far outside the photosphere.  We depict the trajectories of winds
(black lines), propagation of a shock (red line), and position of
the photosphere (blue line), in Figure \ref{hr_mass_velocity_wind}b,
where we assume ballistic motion of wind fluid,
i.e., the velocity of gas is constant outside the photosphere.
Before the optical maximum, the photospheric velocity decreases 
with time, so each locus departs from each other. 
After the optical maximum, on the other hand, 
the trajectories converge, i.e., the wind ejected later is 
catching up the matter previously ejected. 
Thus, matter will be compressed which causes a strong shock wave. 
The mass of shocked shell ($M_{\rm shell}$) is increasing with time and 
reached about 90\% of the total ejecta mass, i.e., $M_{\rm shell}\sim
2.7 \times 10^{-5} ~M_\sun$.  Thus, a large part of nova ejecta is
eventually confined to the shocked shell \citep{hac22k}.



Theoretically, we predict that a shock wave arises after the maximum 
expansion of photosphere and moves outward far outside the photosphere. 
This property is common among all the WD masses and mass-accretion rates   
as far as a main driving force of winds is the radiative opacity.

High energy photons such as hard X-rays could be emitted at the shock.
However, they are not always detected due to absorption by the shocked
shell itself.  We plot the hydrogen column density $N_{\rm H}$ toward 
the shock, i.e., from observers to the shock front, 
in Figure \ref{hr_mass_velocity_wind}d.
In our 1.0 $M_\sun$ WD model, the column density is so high 
($N_{\rm H}\sim 10^{24}$ cm$^{-2}$) in the post-maximum phase,
so the thermal hard X-rays do not escape from the shocked shell. 
Thermal hard X-rays from the shock would be detected in nova outbursts 
only when the column density becomes low.

\section{Light curve model}
\label{light_curve_model}

\subsection{Super Eddington luminosity}
\label{super_eddington}

Figure \ref{hr_mass_velocity_wind}a shows the Eddington limit for 
a $1.0 ~M_\sun$ WD by the short horizontal bar, that is,
\begin{equation}
L_{\rm Edd} = {{4\pi cGM_{\rm WD}} \over {\kappa}},
\label{eddington_luminosity_def}
\end{equation}
where $\kappa$ is the electron scattering opacity of
$\kappa = 0.2(1+X)$ g$^{-1}$ cm$^{2}$.
The photospheric luminosity of our $1.0 ~M_\sun$ evolution 
model does not exceed the Eddington limit. 
The sub-Eddington luminosity is commonly obtained in many 
evolution calculations of novae even in very massive WDs 
\citep[e.g., a 1.38 $M_\sun$ WD model of ][]{kat17sh}.

On the other hand, observed brightnesses
of novae often exceed the Eddington limit \citep[e.g.,][]{del20i}.  
This cannot be explained by our photospheric luminosity (blackbody 
approximation).  In addition, nova spectra sometimes show a flat pattern,
$F_\nu =$ constant against the frequency $\nu$ \citep[e.g.,][soon after 
optical maximum of  V1500 Cyg, one of the brightest novae]{gal76, enn77},
which is different from the blackbody spectrum.
\citet{hac06kb} pointed out that free-free emission 
from nova ejecta outside the photosphere dominates nova
spectra that mainly contributes to the optical luminosity.
They proposed a description formula of free-free flux 
from nova winds as
\begin{eqnarray}
L_{\nu, \rm ff,wind} &=& 4\pi R_{\rm ph}^2 F_\nu 
\propto  \int_{R_{\rm ph}}^{\infty} N_{\rm e} N_{\rm i} d V
\propto \int_{R_{\rm ph}}^{\infty} \rho^2 d V 
\propto \int_{R_{\rm ph}}^{\infty} {{\dot M^2_{\rm wind}} 
\over{v^2_{\rm ph} r^4}} r^2 dr 
 \propto      {{\dot M^2_{\rm wind}} 
\over{v^2_{\rm ph} R_{\rm ph}}},
\label{free-free_flux_definition}
\end{eqnarray}
where $N_{\rm e}$ and $N_{\rm i}$ are the electron number density and
ion number density, respectively, and
we assume a steady-state wind of $\dot{M}_{\rm wind}= 4 \pi r^2 v \rho$
and $\int d V$ and $\int r^2d r$ the volume integration.
We define the $V$ band flux $L_{V, \rm ff}$ as 
\begin{equation}
L_{V, \rm ff,wind} = A_{\rm ff} ~{{\dot M^2_{\rm wind}} 
\over{v^2_{\rm ph} R_{\rm ph}}},
\label{free-free_flux_v-band}
\end{equation}
where the coefficient $A_{\rm ff}$ was determined by 
\citet{hac10k, hac15k, hac16k} and \citet{hac20skhs} for various 
sets of WD mass and chemical composition. 
As the free-free emission flux increases with the wind mass-loss rate, 
it takes the maximum value at the maximum expansion of photosphere.
It decreases in the decay phase because the wind mass-loss rate decreases
with time.  The physical meaning of this formulation is described in more
detail in \citet{hac20skhs}. 
The total $V$ band flux is the summation of the free-free emission luminosity
and the $V$ band flux of the photospheric luminosity $L_{\rm ph}$, 
\begin{equation}
L_{V, \rm total} = L_{V, \rm ff,wind} + L_{V, \rm ph}.
\label{luminosity_summation_flux_v-band}
\end{equation}


\begin{figure*}
\gridline{\fig{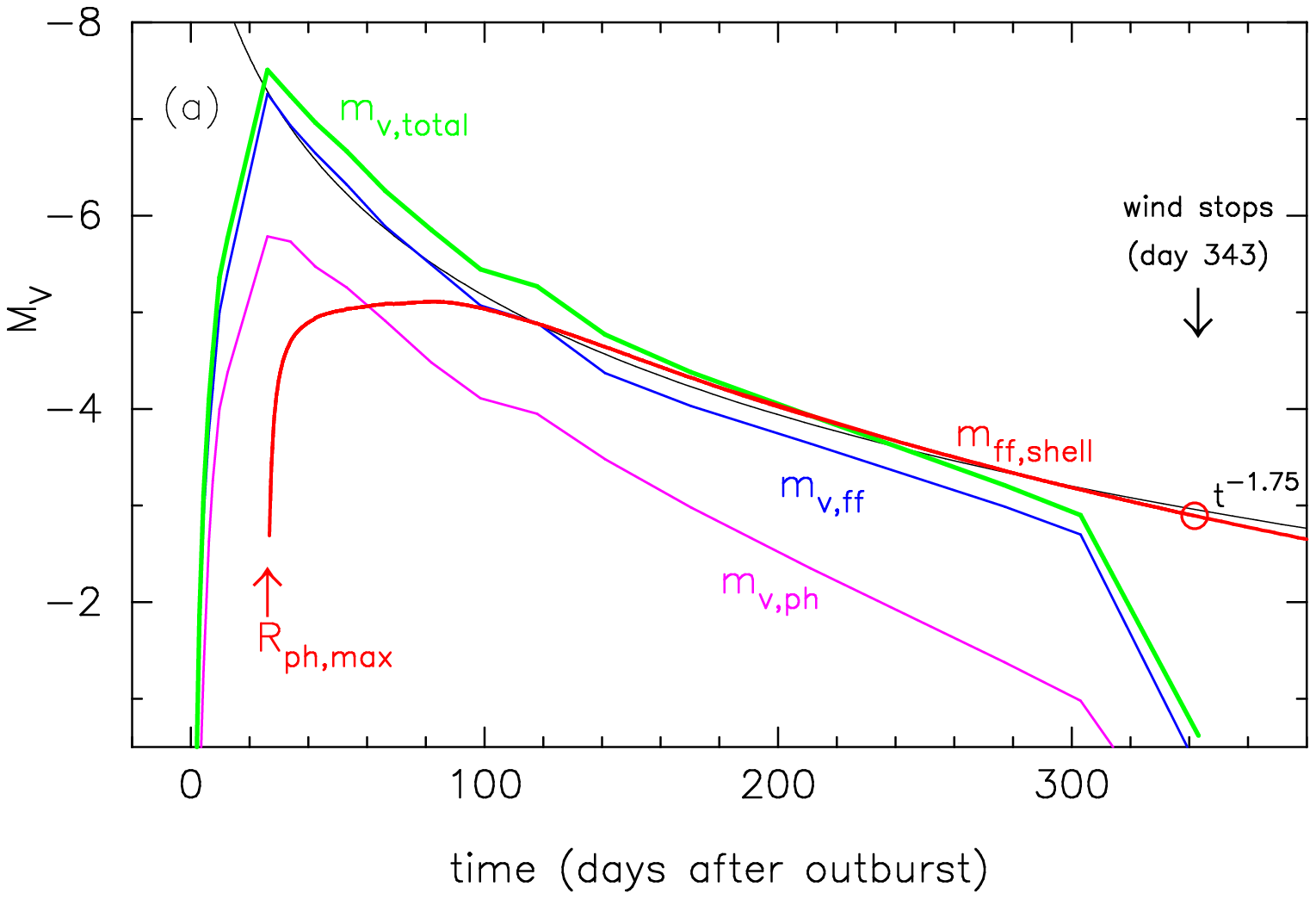}{0.5\textwidth}{(a)}
          \fig{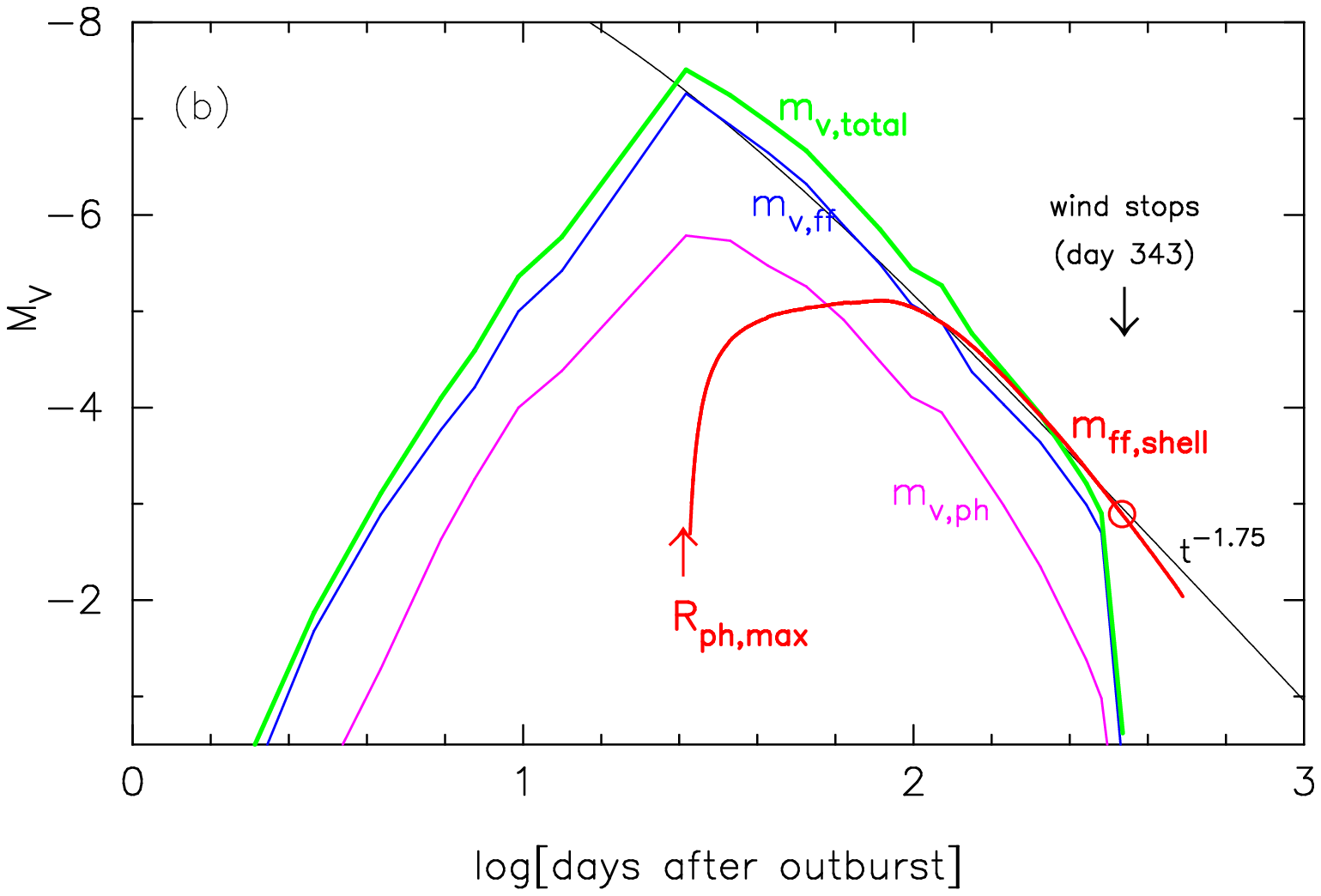}{0.5\textwidth}{(b)}
          }
\epsscale{1.15}
\caption{
The $V$ light curves for the self-consistent nova model,
a $1.0~M_\sun$ WD with a mass-accretion rate of $5\times 10^{-9} ~M_\sun$
yr$^{-1}$.  Three fluxes are plotted: photospheric emission (magenta
line labeled $m_{V, \rm ph}$), free-free emission (blue line, 
$m_{V, \rm ff}$), and total of them (green line,
$m_{V, \rm total}$).  The flux of the shell emission is also added
with the red line labeled $m_{\rm ff, shell}$. 
The open red circle on the red line indicates
the epoch when optically thick winds stop.  The right edge of the red line
corresponds to the epoch when the shock disappears.
We also plot the $V$ magnitudes of equation 
(\ref{nebular_emission_brightness}) by the thin black line
labeled $t^{-1.75}$, and indicate two epochs, one is at maximum expansion
of the photosphere on day 26 and the other is that optically thick winds
stop 343 days after the outburst.
{\bf (a)} Each brightness in the absolute $V$ magnitude
is plotted against a linear timescale (days after outburst)
{\bf (b)} Same as in panel (a), but in a logarithmic timescale 
($\log$[days after outburst]).
\label{optical_m10selfcon_abs_magnitudes}}
\end{figure*}

Figure \ref{optical_m10selfcon_abs_magnitudes} shows 
the $V$ light curves of the 1.0 $M_\sun$ WD described in 
Appendix \ref{section_nova_model}.
Here, $m_{V, \rm ph}$ is calculated from the photospheric temperature and
luminosity with the blackbody assumption and the canonical $V$ band filter. 
At the maximum expansion of photosphere, the temperature decreases down 
to $T_{\rm ph}\sim 8000$ K (Figure \ref{hr_mass_velocity_wind}a),
so that the absolute $V$ magnitude reaches $M_V\sim -6$ at the optical
$V$ peak.  It is interesting that three light curves of $m_{V, \rm total}$, 
$m_{V, \rm ff}$, and $m_{V, \rm ph}$, which correspond to the $V$ magnitudes
of $L_{V, \rm total}$, $ L_{V, \rm ff,wind}$, and $L_{V, \rm ph}$,
respectively, have a very similar shape, 
but the free-free $V$ ($M_V \sim -7.2$) and total $V$ ($M_V \sim -7.5$)
magnitudes are much brighter than the Eddington limit ($M_V \sim -6$).



\begin{figure*}
\gridline{\fig{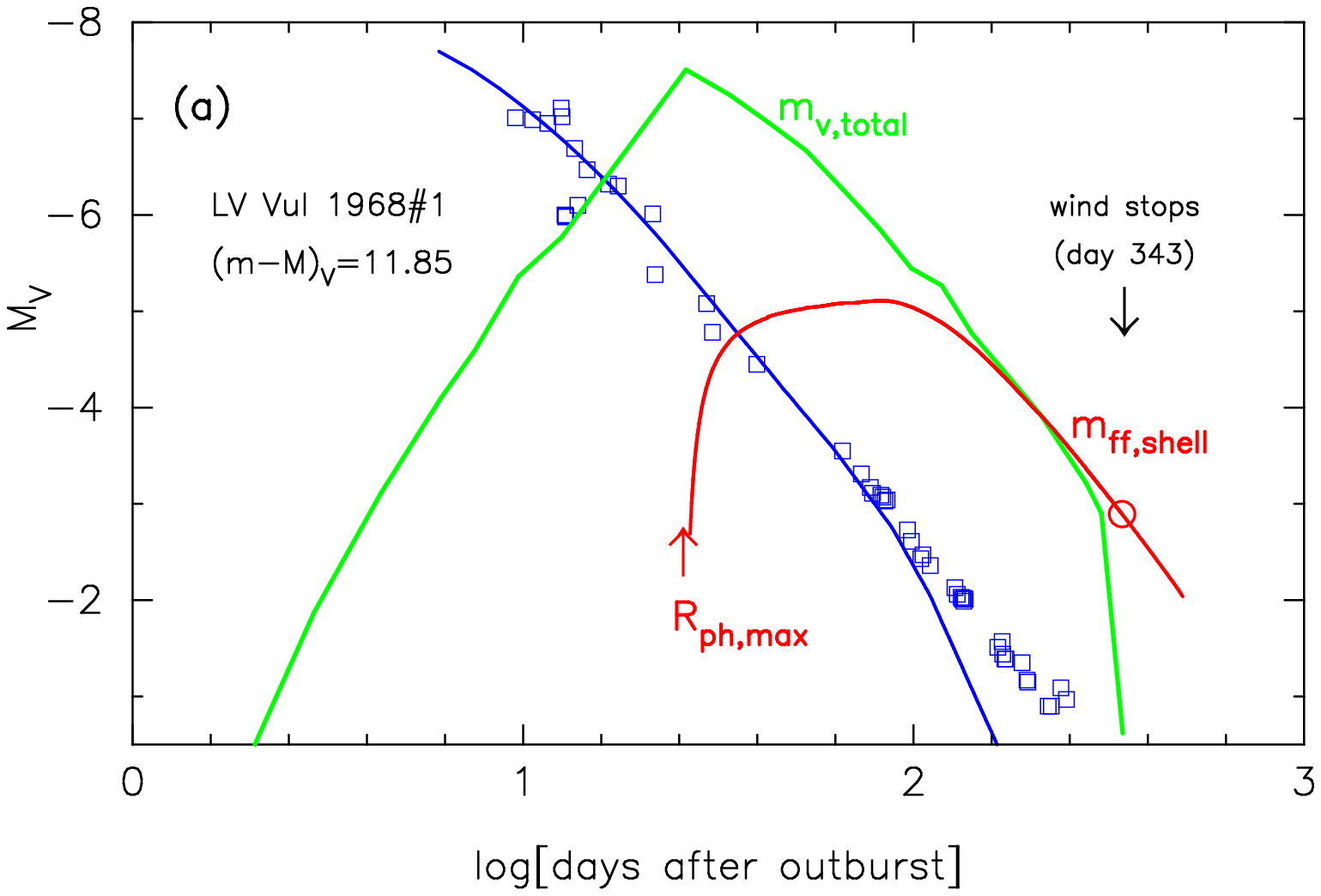}{0.5\textwidth}{(a)}
          \fig{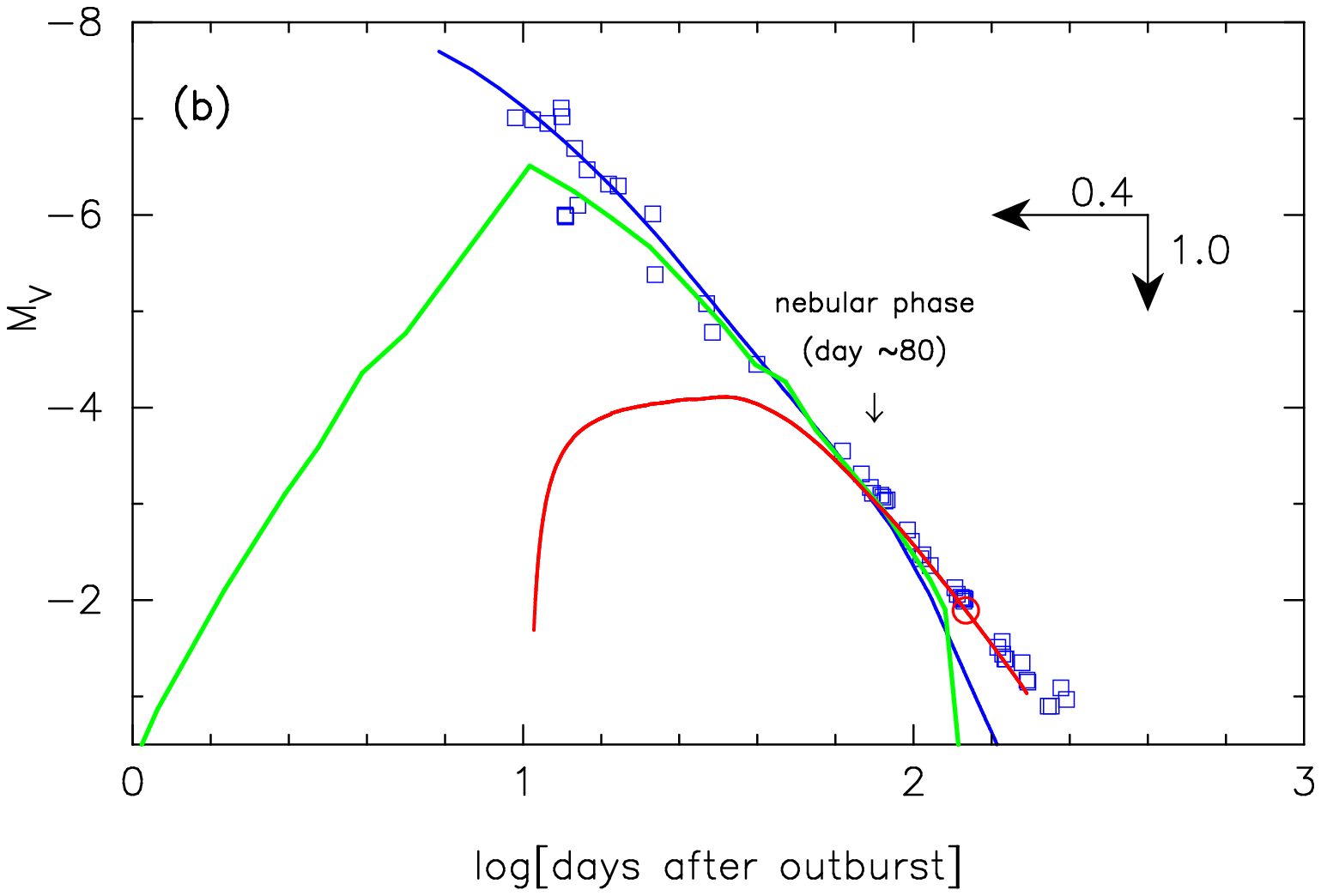}{0.5\textwidth}{(b)}
          }
\epsscale{1.15}
\caption{
{\bf (a) before time-stretching:} 
The $V$ light curves for the classical nova LV Vul 1968\#1
(open blue squares) and our standard 1.0 $M_\sun$ WD model 
(green and red lines, both of which are the same as those in Figure
\ref{optical_m10selfcon_abs_magnitudes}b).
The data of LV Vul are the same as those 
in \citet{hac16kb}.  We add a $0.98 ~M_\sun$ WD (CO3) model light
curve (blue line) that is the best fit model to LV Vul \citep{hac16k}.
The timescale of LV Vul is about 2.5 times shorter than that of
the 1.0 $M_\sun$ nova model. 
{\bf (b) after time-stretching:} Light curve of LV Vul 
compared with the time-stretched 1.0 $M_\sun$ model 
(using equation (\ref{time-stretching_general})). 
Both the green and red lines move leftward by $\Delta \log t = -0.4$
and downward by $\Delta M_V = 1.0$. 
\label{optical_m10_abs_lv_vul_mag}}
\end{figure*}


\subsection{Light curve model in the nebular phase}
\label{light_curve_nebular_phase}

The $V$ brightness of free-free emission in Figure  
\ref{optical_m10selfcon_abs_magnitudes} rapidly decreases just before
the optically thick winds stop. 
On the other hand, the shocked shell emission keeps brightness
by an extra time of $t_{\rm ret}$ until the shock disappears
(Section \ref{shock_duration_hard_x-ray}). 
Thus, the contribution from the shocked shell is important 
in the later decay phase. 
We formulate the free-free flux from the shocked shell by
\begin{eqnarray}
L_{\nu, \rm ff,shell} & = & 4\pi R_{\rm shock}^2 F_{\nu, \rm ff,shell} 
 \propto  \int^{\rm shell} N_{\rm e} N_{\rm i} d V
\propto \int^{\rm shell} \rho^2 d V 
 \propto  {M_{\rm shell}^2 
\over {4\pi R_{\rm shock}^2 h}},
\label{flux_shocked_shell_ff_flux}
\end{eqnarray}
where $N_{\rm e}$, $N_{\rm i}$, $M_{\rm shell}$, $R_{\rm shock}$,
and $h$ are the electron number density, ion number density, mass of
the shocked shell, radius of the shocked shell, and the thickness
of the shocked shell, respectively, and $\int dV$ means the volume 
integration in the shocked shell. We have
$M_{\rm shell}= \rho 4\pi R_{\rm shock}^2 h = \rho V_{\rm shell}$
for a geometrically thin shocked shell.
The bound-free and bound-bound emissions also depend on the square
of the density \citep[e.g.,][]{wil92}, i.e., $\propto \rho^2$, 
so that the total luminosity from the shocked shell is estimated from
the same dependency as that of free-free emission,
\begin{equation}
L_{V, \rm shell} = B_{\rm ff} 
{{M_{\rm shell}^2} \over {4\pi R_{\rm shock}^2 h}}.
\label{luminosity_shocked_shell_ff_flux}
\end{equation}
We plot the temporal variation of this luminosity in Figure
\ref{optical_m10selfcon_abs_magnitudes} by the red line.
Here, we assume that the thickness of the shocked shell is very small
(very thin) and does not change in time, i.e., $h=$constant in time.
The proportionality constant ($B_{\rm ff}$ or) $B_{\rm ff}/h$ is
determined to smoothly connect to the $m_{V, \rm total}$ 
(green line labeled $m_{V, \rm total}$ in Figure 
\ref{optical_m10selfcon_abs_magnitudes}).
We calculate $L_{V, \rm shell}$, through the epoch when
the optically thick winds stop, which is denoted by an open red circle,
until the shock disappears, the right edge of the red line.
We added the $V$ light curve of equation (\ref{nebular_emission_brightness})
with $(m-M)_V=0.0$ and $\Delta V= -2.4$ by the black line labeled $t^{-1.75}$,
which is a decline trend of YZ Ret in the nebular phase.

Now we see that the three curves in 
Figure \ref{optical_m10selfcon_abs_magnitudes} 
show the same decline trend of $t^{-1.75}$ from day 200 to 
day 300.
The three curves are 
(i) $L_{V, \rm total}$ in equation (\ref{luminosity_summation_flux_v-band}),
that is, the free-free emission from the optically thick
winds ($L_{V, \rm ff}$ in equation (\ref{free-free_flux_v-band})) plus
photospheric emission ($L_{V, \rm ph}$), 
(ii) emission from the shocked shell ($L_{V, \rm shell}$ in equation 
(\ref{luminosity_shocked_shell_ff_flux})), and 
(iii) $V$ light curve of YZ Ret 
(equation (\ref{nebular_emission_brightness})). 
\citet{hac06kb} called the decline trend of $L_V \propto t^{-1.75}$ 
the universal decline law of classical novae based on the 
free-free emission from the optically thick winds. 
We conclude that this universal decline law can be extended to the nebular
phase, in which free-free, bound-free, or bound-bound emission from
the shocked shell becomes dominant.

In the present paper, we separately treat the two luminosities of
$L_{V, \rm total}= L_{V, \rm ff,wind} + L_{V, \rm ph}$
(equation (\ref{luminosity_summation_flux_v-band})) and
$L_{V, \rm shell}$ (equation (\ref{luminosity_shocked_shell_ff_flux})). 
When we fit observational data of each nova, we assume 
the total flux is
\begin{equation}
L_V = \max\left( L_{V, \rm ff,wind}+L_{V, \rm ph}, 
~L_{V, \rm shell}\right),
\label{final_luminosity_wind_shell}
\end{equation}
instead of simple summation of the two, 
because the shocked shell may absorb a part 
of $L_{V, \rm ff,wind}+L_{V, \rm ph}$ and reemits
as a part of $L_{V, \rm shell}$.  This assumption could be supported
by the smooth, almost straight decline without a bump in the LV Vul 
light curve (Figure \ref{optical_m10_abs_lv_vul_mag}) or other novae in 
previous sections, although we need radiative transfer calculation
to finally fix this assumption.

\subsection{Time-stretching method in LV Vul}
\label{time_stretchin_method}

\citet{hac10k} found that two different model light curves,
e.g., different WD mass or speed class,
overlap each other if the timescale of one of them is squeezed by
a factor of $f_{\rm s}$, i.e., as $t/f_{\rm s}$.
The normalization factor is $f_{\rm s}< 1$ for a faster nova
(corresponding to a more massive WD), and $f_{\rm s}> 1$ for a slower nova
(a less massive WD).  The absolute $V$ brightnesses is normalized to be
$M_V-2.5\log f_{\rm s}$.  These two light curves
overlap each other in the $(t/f_{\rm s})$-$(M_V-2.5\log f_{\rm s})$
plane \citep[see, e.g., Figures 48 and 49 of][]{hac18kb}.
\citet{hac19kb} reformulated this property:
if the $V$ light curve of a template nova (time $t$) overlaps with
that of a target nova (time $t'=t/f_{\rm s}$), we have the relation
\begin{eqnarray}
\left( M_V[t] \right)_{\rm template}
&=& \left( M'_V[t'] \right)_{\rm target} \cr
&=& \left( M_V[t/f_{\rm s}]-2.5\log f_{\rm s} \right)_{\rm target},
\label{time-stretching_general}
\end{eqnarray}
where $M_V[t]$ is the original absolute $V$ brightness and $M'_V[t']$
is the time-normalized brightness after time-normalization
of $t'=t/f_{\rm s}$.
This property was applied to many novae and their $f_{\rm s}$
are determined \citep[e.g.,][]{hac16k, hac18kb, hac19ka,
hac21k, hac20skhs}.

Now we apply this method to a pair of LV Vul and our 1.0 $M_\sun$ WD model.   
Figure \ref{optical_m10_abs_lv_vul_mag} shows the $V$ band magnitudes
(open blue squares) of LV Vul.
We set the outburst day of LV Vul as JD 2,439,962.0 
(UT 1968 April 15.5).
The best fit model for LV Vul is a $0.98 ~M_\sun$ WD (CO3)
\citep{hac16k}, the light curve of which is the summation of free-free
plus blackbody emissions calculated from equations 
(\ref{free-free_flux_v-band}) and (\ref{luminosity_summation_flux_v-band})
and shown by the blue line.  This blue line well follows the LV Vul $V$ data 
from day $\sim 10$ to day $\sim 100$.  

Figure \ref{optical_m10_abs_lv_vul_mag}a also shows the light curve 
of our 1.0 $M_\sun$ model.  The decay timescale 
is about 2.5 times longer than that of LV Vul ($f_s=2.5$). 
We apply the stretching method to this model 
and move it leftward by $- \log f_{\rm s} = - \log 2.5 = -0.4$
and downward by $2.5\log f_{\rm s}= 1.0$.  As shown in Figure 
\ref{optical_m10_abs_lv_vul_mag}b, 
the green line well overlaps with the blue line. 
We also apply this method to the shocked shell (red line)  
and confirm that the red line overlaps with the $V$ data (open blue squares)
from day $\sim 80$ to day $\sim 150$.


Figure 6 of \citet{hac16kb} shows the LV Vul data in  
the $(B-V)_0$-$M_V$ color-magnitude diagram in which the light curve 
shows a split to two branches at $M_V=-3.3$ between  
\citet{gry69}'s and \citet{fer69}'s data.  Here $(B-V)_0$ is the intrinsic
$B-V$ color.  This suggests that 
a transition to the nebular phase occurred on $\sim 80$ days after
the outburst (see Figure \ref{optical_m10_abs_lv_vul_mag}b).
The 1968 July 21 spectrum
of LV Vul clearly shows a nebular phase spectrum \citep{hut70}.  Thus, the
nova had entered the nebular phase, at least, 97 days after the outburst.
The previous observation on 1968 July 1, 77 days after the outburst,
did not show strong nebular lines yet. 

The red line crosses the blue line on day $\sim 80$ in Figure 
\ref{optical_m10_abs_lv_vul_mag}b.  This suggests that the shell emission
dominates the $V$ flux of the nova after day $\sim 80$.
Thus, we may conclude that the $t^{-1.75}$ decline law holds
even in the nebular phase (after day $\sim 80$).

\clearpage








\begin{thebibliography}{}


\bibitem[Ackermann et al. (2014)]{ack14}
Ackermann, M., Ajello, M., Albert, A., et al. 2014, Science, 345, 554,
\doi{10.1126/science.1253947}



\bibitem[Aydi et al. (2020a)]{ayd20ci}
Aydi, E., Chomiuk, L., Izzo, L., et al. 2020a, \apj, 905, 62, 
\doi{10.3847/1538-4357/abc3bb}



\bibitem[Aydi et al. (2020b)]{ayd20bc}
Aydi, E., Buckley, D. A. H., Chomiuk, L., et al. 2020b, ATel, 13867, 1



\bibitem[Bailer-Jones et al. (2021)]{bai21rf}
Bailer-Jones, C. A. L., Rybizki, J., Fouesneau, M., Demleitner, M.,
\& Andrae, R. 2021, \aj, 161, 147,
\doi{10.3847/1538-3881/abd806}

\bibitem[Beals (1931)]{bea31}
Beals, C. S. 1931, \mnras, 91, 966, \doi{10.1093/mnras/91.9.966}

\bibitem[Bertout \& Magnan (1987)]{ber87m}
Bertout, C., \& Magnan, C. 1987, \aap, 183, 319





%



%










\bibitem[Chomiuk et al. (2021)]{cho21ms}
Chomiuk, L., Metzger, B. D., \& Shen, K. J. 2021, 
Annual Review of Astronomy and Astrophysics, 59, 48,
\doi{10.1146/annurev-astro-112420-114502} 



\bibitem[della Valle \& Izzo (2020)]{del20i}
della Valle, M., \& Izzo, L. 2020, The Astronomy and Astrophysics Review,
28, 3, \doi{10.1007/s00159-020-0124-6}

%










\bibitem[Ennis et al. (1977)]{enn77}
Ennis, D., Becklin, E. E., Beckwith, S., et al. 1977, \apj, 214, 478,
\doi{10.1086/155273}


\bibitem[Fernie (1969)]{fer69}
Fernie, J. D. 1969, \pasp, 81, 374, \doi{10.1086/128790}




\bibitem[Fujikawa et al. (2012)]{fuj12}
Fujikawa, S., Yamaoka, H., Nakano, S. 2012, CBET, 3202, 1

\bibitem[Galan \& Mikolajewska (2020)]{gal20m}
Galan, C. \& Mikolajewska, J. 2020, ATel, 14149, 1

\bibitem[Gallagher \& Ney (1976)]{gal76}
Gallagher, J. S., \& Ney, E. P. 1976, \apj, 204, L35,
\doi{10.1086/182049}




\bibitem[Grygar (1969)]{gry69}
Grygar, J. 1969, Inf.\ Bull.\ Variable Stars, 371, 1
 


\bibitem[Hachisu \& Kato (2006)]{hac06kb}
Hachisu, I., \& Kato, M. 2006, \apjs, 167, 59,
\doi{10.1086/508063}


\bibitem[Hachisu \& Kato (2010)]{hac10k}
Hachisu, I., \& Kato, M. 2010, \apj, 709, 680,
\doi{10.1088/0004-637X/709/2/680}

\bibitem[Hachisu \& Kato (2014)]{hac14k}
Hachisu, I., \& Kato, M. 2014, \apj, 785, 97,
\doi{10.1088/0004-637X/785/2/97}

\bibitem[Hachisu \& Kato (2015)]{hac15k}
Hachisu, I., \& Kato, M. 2015, \apj, 798, 76,
\doi{10.1088/0004-637X/798/2/76}

\bibitem[Hachisu \& Kato (2016a)]{hac16k}
Hachisu, I., \& Kato, M. 2016a, \apj, 816, 26,
\doi{10.3847/0004-637X/816/1/26}
%

\bibitem[Hachisu \& Kato (2016b)]{hac16kb}
Hachisu, I., \& Kato, M. 2016b, \apjs, 223, 21, 
\doi{10.3847/0067-0049/223/2/21}
%


\bibitem[Hachisu \& Kato (2018a)]{hac18k}
Hachisu, I., \& Kato, M. 2018a, \apj, 858, 108, \doi{10.3847/1538-4357/aabee0}

\bibitem[Hachisu \& Kato (2018b)]{hac18kb}
Hachisu, I., \& Kato, M. 2018b, \apjs, 237, 4, \doi{10.3847/1538-4365/aac833}

\bibitem[Hachisu \& Kato (2019a)]{hac19ka}
Hachisu, I., \& Kato, M. 2019a, \apjs, 241, 4, \doi{10.3847/1538-4365/ab0202}

\bibitem[Hachisu \& Kato (2019b)]{hac19kb}
Hachisu, I., \& Kato, M. 2019b, \apjs, 242, 18, \doi{10.3847/1538-4365/ab1b43}

\bibitem[Hachisu \& Kato (2021)]{hac21k}
Hachisu, I., \& Kato, M. 2021, \apjs, 253, 27, \doi{10.3847/1538-4365/abd31e}

\bibitem[Hachisu \& Kato (2022)]{hac22k}
Hachisu, I., \& Kato, M. 2022, \apj, 939, 1, \doi{10.3847/1538-4357/ac9475}



\bibitem[Hachisu et al. (2020)]{hac20skhs}
Hachisu, I., Saio, H., Kato, M., Henze, M. \& Shafter, A.W.  2020, 
\apj, 902, 91, \doi{10.3847/1538-4357/abb5fa}

\bibitem[Hutchings (1972)]{hut72}
Hutchings, J. B. 1972, \mnras, 158, 177, \doi{10.1093/mnras/158.2.177}




\bibitem[Hutchings (1970)]{hut70}
Hutchings, J. B. 1970, \pasp, 82, 603, \doi{10.1086/198237}

\bibitem[Iglesias \& Rogers (1996)]{igl96r}
Iglesias, C. A., \& Rogers, F. J. 1996, \apj, 464, 943, \doi{10.1086/177381}


\bibitem[Izzo et al. (2020)]{izz20ma}
Izzo, L., Molaro, P., Aydi, E., et al. 2020, The Astronomer's Telegram,
14048, 1

\bibitem[Jayasinghe et al. (2019)]{jay19sk}
Jayasinghe, T., Stanek, K. Z., Kochanek, C. S., et al. 2019, \mnras, 
486, 1907, \doi{10.1093/mnras/stz844}



\bibitem[Kato \& Hachisu (1994)]{kat94h}
Kato, M., \& Hachisu, I., 1994, \apj, 437, 802, \doi{10.1086/175041}







\bibitem[Kato et al. (2017)]{kat17sh}
Kato, M., Saio, H., \& Hachisu, I., 2017, \apj, 838, 153,
\doi{10.3847/1538-4357/838/2/153} 

\bibitem[Kato et al. (2021)]{kat21sh}
Kato, M., Saio, H., \& Hachisu, I., 2021, \pasj, 73, 1137,
\doi{10.1093/pasj/psab064}

\bibitem[Kato et al. (2022a)]{kat22sha}
Kato, M., Saio, H., \& Hachisu, I. 2022a, \pasj, 74, 1005, 
\doi{10.1093/pasj/psac051}

\bibitem[Kato et al. (2022b)]{kat22shb}
Kato, M., Saio, H., \& Hachisu, I. 2022b, \apjl, 935, L15, 
\doi{10.3847/2041-8213/ac85c1}


\bibitem[Kato et al. (2022c)]{kat22shc}
Kato, M., Saio, H., \& Hachisu, I. 2022c, RNAAS, 6, 258,
\doi{10.3847/2515-5172/aca8af}



\bibitem[K{\"o}nig et al. (2022)]{kon22wa}
K{\"o}nig, O., Wilms, J., Arcodia, R., et al. 2022, \nat, 605, 248,
\doi{10.1038/s41586-022-04635-y}








\bibitem[Lockwood \& Millis (1976)]{loc76m}
Lockwood, G. W., \& Millis, R. L. 1976, \pasp, 88, 235, 
\doi{10.1086/129935}



\bibitem[Martin et al. (2018)]{mar18dj}
Martin, P., Dubus, G., Jean, P., Tatischeff, V., \& Dosne, C. 2018, \aap,
612, A38, \doi{10.1051/0004-6361/201731692}



\bibitem[McLaughlin (1942)]{mcl42}
McLaughlin, D. B. 1942, \apj, 95, 428, \doi{10.1086/144414} 

\bibitem[Mclaughlin (1943)]{mcl43}
McLaughlin, D. B. 1943, Publications of the Observatory of the University
of Michigan, 8, 149 







\bibitem[McNaught (2020)]{mcn20}
McNaught, R. H. 2020, CBET, 4812, 2

\bibitem[McLoughlin et al. (2021)]{mcl21bl}
McLoughlin, D., Blundell, K. M., Lee, S., \& McCowage, C. 2021, \mnras,
503, 704, \doi{10.1093/mnras/stab581}

\bibitem[Metzger et al. (2014)]{met14hv}
Metzger, B. D., Hasco\"et, R., Vurm, I., et al. 2014, \mnras, 442, 713,
\doi{10.1093.mnras.stu844}

\bibitem[Metzger et al. (2015)]{met15fv}
Metzger, B. D., Finzell, T., Vurm, I., et al. 2015, \mnras, 450, 2739,
\doi{10.1093/mnras/stv742}








\bibitem[Munari et al. (2013)]{mun13dc}
Munari, U., Dallaporta, S., Castellani, F., et al. 2013, \mnras, 435, 771,
\doi{10.1093/mnras/stt1340}

%


\bibitem[Murawski (2019)]{mur19}
Murawski, G. 2019, Astronomical Report, IX, 33



\bibitem[Ness et al. (2013)]{nes13oh}
Ness, J. -U., Osborne, J. P., Henze, M., et al. 2013, \aap, 559, A50,
\doi{10.1051/0004-6361/201322415}






\bibitem[Orio et al. (2022)]{ori22gg}
Orio, M., Gendreau, K., Giese, M., et al. 2022, \apj, 932, 45,
\doi{10.3847/1538-4357/ac63be}

\bibitem[Osaki (1996)]{osa96}
Osaki, Y. 1996, \pasp, 108, 39, \doi{10.1086/133689}






\bibitem[Payne-Gaposchkin (1957)]{pay57}
Payne-Gaposchkin, C. 1957, The Galactic Novae (Amsterdam: North-Holland)

\bibitem[Pei et al. (2020)]{pei20og}
Pei, S., Orio, M., Gendreau, K., et al. 2020,
The Astronomer's Telegram, 14067, 1





 


\bibitem[Raj et al. (2012)]{raj12ab}
Raj, A., Ashok, N. M., Banerjee, D. P. K., et al.
2012, \mnras, 425, 2576, \doi{10.1111/j.1365-2966.2012.21739.x}


\bibitem[Schaefer (2022)]{schaefer22}
Schaefer, B.E., 2022,\mnras, 517, 3640, \doi{10.1093/mnras/stac2089}





\bibitem[Shappee et al. (2014)]{sha14pg}
Shappee , B. J., Prieto, J. L., Grupe, D., et al. 2014, \apj, 788, 48,
\doi{10.1088/0004-637X/788/1/48}


\bibitem[Shara et al. (2012)]{sha12zd}
Shara, M. M., Zurek, D., De Marco, O., et al. 2012, \aj, 143, 143,
\doi{10.1088/0004-6256/143/6/143}



\bibitem[Sitko et al. (2020)]{sit20rr}
Sitko, M. L., Rudy, R. J., \& Russell, R. W., 2020, ATel, 14205, 1 



\bibitem[Slavin et al. (1995)]{sla95od}
Slavin, A. J., O'Brien, T. J., Dunlop, J. S. 1995, \mnras, 276, 353,
\doi{10.1093/mnras/276.2.353}


\bibitem[Sokolovsky et al. (2020a)]{sok20ac1}
Sokolovsky, K. V., Aydi, E., Chomiuk, L., et al. 2020, ATel, 13900, 1 

\bibitem[Sokolovsky et al. (2020b)]{sok20ac2}
Sokolovsky, K. V., Aydi, E., Chomiuk, L., et al. 2020, ATel, 14043, 1 


\bibitem[Sokolovsky et al. (2022)]{sok22ll}
Sokolovsky, K. V., Li, K.-L., \& Lopes de Oliveira, R. 2022, \mnras, 
514, 2239, \doi{10.1093/mnras/stac1440}










\bibitem[Tarasova (2016)]{tar16}
Tarasova, T. N., 2016, Astronomy Reports, 60, 1052, 
\doi{10.1134/S106377291611007X}




\bibitem[Wagenblast et al. (1983)]{wag83bb}
Wagenblast, R., Bertout, C., \& Bastian, U. 1983, \aap, 120, 6


\bibitem[Williams (1992)]{wil92}
Williams, R. 1992, \aj, 104, 725, \doi{10.1086/116268}






\bibitem[Woodward et al. (1997)]{woo97gj}
Woodward, C. E., Gehrz, R. D., Jones, T. J., Lawrence, G. F.,
\& Skrutskie, M. F. 1997, \apj, 477, 817, \doi{10.1086/303739}


\end{thebibliography}
\end{document}